\documentclass[numberedappendix]{emulateapj}

\usepackage{multirow}
\usepackage{graphicx}
\usepackage{natbib}
\usepackage{amsmath}
\usepackage{bm}

\newcommand{\vecv}{\mathbf{v}}
\newcommand{\vecu}{\mathbf{u}}
\newcommand{\vecx}{\mathbf{x}}
\newcommand{\vecxi}{{\bm \xi}}
\newcommand{\ts}{T_*}
\newcommand{\cs}{c_{s,*}}
\newcommand{\hs}{h_*}
\newcommand{\rhos}{\rho_*}
\newcommand{\fE}{f_{\rm E,*}}
\newcommand{\fEl}{f_{\rm E}}
\newcommand{\fEc}{f_{\rm E,crit}}
\newcommand{\mh}{m_{\rm H}}
\newcommand{\msun}{M_{\odot}}
\newcommand{\lsun}{L_{\odot}}
\newcommand{\fzavg}{\langle f_{{\rm rad},z}\rangle}
\newcommand{\fEavg}{\langle f_{\rm E}\rangle}
\newcommand{\fEdir}{\langle f_{\rm E,dir}\rangle}
\newcommand{\ft}{f_{\rm trap}}

\newcommand{\jcam}{J.~Comp.~App.~Math.}
\newcommand{\jcompphys}{J.~Comp.~Phys.}

\newcommand{\red}[1]{#1}

\begin{document}

\title{Direct Numerical Simulation of Radiation Pressure-Driven Turbulence and Winds in Star Clusters and Galactic Disks}

\author{Mark R. Krumholz}
\affil{Department of Astronomy \& Astrophysics, University of California, Santa 
Cruz, CA 95064 USA}
\email{krumholz@ucolick.org}

\author{Todd A. Thompson}
\affil{Department of Astronomy and Center for Cosmology \& Astro-Particle Physics \\The Ohio State University, Columbus, OH 43210-1173 USA}
\email{thompson@astronomy.ohio-state.edu}

\slugcomment{Accepted for publication in ApJ}

\begin{abstract}
The pressure exerted by the radiation of young stars may be an important feedback mechanism that drives turbulence and winds in forming star clusters and the disks of starburst galaxies. However, there is great uncertainty in how efficiently radiation couples to matter in these high optical depth environments.  In particular, it is unclear what levels of turbulence the radiation can produce, and whether the infrared radiation trapped by the dust opacity can give rise to heavily mass-loaded winds. In this paper we report a series of 
\red{two-dimensional flux-limited diffusion radiation-hydrodynamics calculations}
performed with the 
code \textsc{orion} in which we drive strong radiation fluxes through columns of dusty matter confined by gravity in order to answer these questions.  We consider both systems where the radiation flux is sub-Eddington throughout the gas column, and those where it is super-Eddington at the midplane but sub-Eddington in the atmosphere. In the latter, we find that the radiation-matter interaction gives rise to radiation-driven Rayleigh-Taylor instability, which drives supersonic turbulence at a level sufficient to fully explain the turbulence seen in Galactic protocluster gas clouds, and to make a non-trivial contribution to the turbulence observed in starburst galaxy disks. However, the instability also produces a channel structure in which the radiation-matter interaction is reduced compared to time-steady analytic models because the radiation field is not fully trapped.  \red{For astrophysical parameters relevant to forming star clusters and starburst galaxies, we find that} this effect reduces the net momentum deposition rate in the dusty gas \red{by a factor of $\sim 2-6$ compared to simple analytic estimates}, and that in steady state the Eddington ratio reaches unity and there are no strong winds.
We provide an approximation formula, appropriate for implementation in analytic models and non-radiative simulations, for the force exerted by the infrared radiation field in this regime.
\end{abstract}


\section{Introduction}
\label{sec:intro}

The idea that the pressure exerted by stellar radiation might have significant dynamical effects on the gas in galaxies is an old one \citep{odell67a, chiao72a, elmegreen83a, ferrara93a, scoville01a, scoville03a}, but it has received significant renewed attention in recent years as a possible explanation for various phenomena in star clusters and galaxies. On galactic scales, \citet{thompson05a} and \citet{murray05a} propose that the force exerted by radiation from newly-formed stars both sets an upper limit on the star formation rate in ultraluminous infrared galaxies (ULIRGs) and drives the highly supersonic motions that are ubiquitously observed in these systems. \citet{andrews11a} extend this analysis and propose that radiation pressure limits on the star formation rate even in normal galaxies. \citet{murray05a} and \citet{murray11a} argue that radiation pressure is also responsible for launching galactic winds (see also Zhang \& Thompson 2012). On subgalactic scales, \citet{krumholz09d} and \citet{fall10a} argue that radiation pressure is the dominant feedback mechanism for massive young star clusters, and that winds driven by radiation momentum set the star formation efficiency in clusters and the cluster mass function.  \citet{murray10a} also argue that radiation pressure is the primary feedback mechanism in massive star clusters, and that it is responsible for limiting the star formation efficiency of giant molecular clouds across a wide range of galactic environments at low- and high-redshift.

This renewed theoretical attention has given rise to a number of approximate implementations of radiation pressure feedback in simulations of galaxy evolution. The earliest of these is the ``momentum-driven wind" model of \citet{oppenheimer06a}, in which they inject momentum into star-forming gas in their cosmological simulations at a rate that depends on both to the star formation rate and the depth of the galactic potential well; the latter dependence is an attempt to approximate the increase in radiative momentum imparted to gas that occurs when the optical depth is high, so every photon is absorbed and reemitted multiple times. More recently, \citet{hopkins11d, hopkins11a, hopkins12a, hopkins12b} have implemented a more sophisticated model for radiation feedback into their isolated galaxy simulations. In their approach, the code identifies contiguous star-forming clumps and then applies an outward radiation force to the gas in them. The force is proportional to the product of the luminosity produced by all stars in the clumps and the gas column density; as with the momentum-driven wind model, the latter dependence is an attempt to capture the effects of force amplification due to radiative trapping.

However, both the analytic models and the resulting semi-analytic implementations of feedback in the simulations contain significant uncertainties. At a basic level, it is unclear whether the radiation is actually able to drive winds from gravitationally bound galaxies or gas clumps. The opacity of dusty material varies with temperature as roughly $\kappa \propto T^2$ (at temperatures $\la 150$ K; \citealt{semenov03a}), and the high optical depths of dusty ULIRGs or star-forming clumps ensures that their central temperatures are higher than the temperatures at their edges. For such objects it is often the case that the Eddington ratio (defined as the ratio of the radiative and gravitational forces per unit mass) is larger than unity for the warm material near the center, but less than unity for the cooler material near the edge. In this case it is unclear whether radiation will launch a wind at all.  \citet{thompson05a}, \citet{murray10a}, and \citet{andrews11a} argue that such a configuration will produce a wind if the temperature and thus the Eddington ratio at the center is sufficiently high, while \citet{krumholz09d} argue that a wind will occur only if the Eddington ratio exceeds unity at the edge of the object.

If there is a wind, a second uncertainty is in how much momentum the radiation deposits in the matter. Consider a source of radiation of luminosity $L$, such as a young star, surrounded by dusty gas. If every photon is absorbed once, the radiation will deposit its full momentum flux $L/c$ in the gas. However, if the medium is so optically thick that each photon is absorbed and re-emitted multiple times, the amount of momentum deposited in the matter could be significantly larger. In the limit where every photon is absorbed many times, all the energy of the radiation field is transformed into kinetic energy of the gas, and the momentum transfer approaches $L/v$, where $v$ is the gas characteristic velocity, and the exact prefactor will depend on the gas velocity and density distribution. The two limiting cases of $L/c$ and $L/v$ may be referred to as the momentum-driven and energy-driven limits, respectively, since in the former case the momentum deposited is limited by the momentum of the radiation field, while in the latter case it is limited by the energy of the radiation field. 

Different authors have come to differing conclusions about where between these limits systems will fall. \citet{krumholz09d} argue that the momentum deposited will never exceed a few $L/c$, while \citet{thompson05a}, \citet{murray10a}, and \citet{andrews11a} argue that in optically thick systems it will be $\tau_{\rm IR} L/c$, where $\tau_{\rm IR}$ is an appropriately defined mean IR optical depth, which can be $\gg 1$.\footnote{
Note that $L/c \ll \tau_{\rm IR} L/c \ll L/v$, so models with $\tau_{\rm IR}$ are intermediate between the pure momentum- and energy-driven limits. However, in the literature feedback models with $\tau_{\rm IR}$ are still referred to as momentum-driven to distinguish them from models in which the energy of hot supernova-shocked gas dominates feedback.} \red{Which argument turns out to be correct has important implications for questions like whether radiation pressure is the dominant mechanism for disrupting most molecular clouds \citep{murray10a} or only those forming the most luminous star clusters \citep{fall10a}, and whether giant clumps in $z\sim 2$ galaxies survive for long times \citep{krumholz10b} or are rapidly disrupted by stellar feedback \citep{hopkins11d, genel12a}.}

The strength and nature of radiation-matter coupling is uncertain because the matter distribution in real galaxies and star clusters is highly non-uniform, leading to complex density fields through which the radiation must pass. Moreover, there are numerous instabilities that can occur when radiation exerts strong forces on matter, such as the photon bubble instability \citep{blaes03a} and the radiation Rayleigh-Taylor instability \citep{krumholz09c, jacquet11a}, and these alter the distribution of both matter and radiation. The question of radiation-matter coupling therefore requires modeling the fully non-linear development of radiation-hydrodynamic instabilities, which in turn requires simulations capable of treating both the radiation field and the matter. Both calculations of radiative transfer through fixed density fields and calculations of the density field in which the radiation field is taken to be uniform (as is assumed, for example, in the \citet{hopkins11a} feedback prescription) are inadequate to the task. To date no simulations capable of answering this question have been reported.

In this paper we explore a simple model system that nonetheless contains many of the essential features required to study non-linear matter-radiation coupling. We use this system to understand the nature of this coupling, and to derive estimates for current unknowns such as the ability of radiation pressure to drive turbulence and winds, and the efficiency with which radiation deposits momentum in the gas. In Section~\ref{sec:model} we describe our model system and the equations that govern it, obtain its important dimensionless numbers, and determine under what circumstances it has an equilibrium state. In Section~\ref{sec:simulations} we describe the numerical methods we use to simulate our model system, and in Section~\ref{sec:simresults} we describe the results of our numerical simulations, the limitations of our calculations, and caveats to our conclusions. Finally, in Section~\ref{sec:discussion} we discuss the implications of our results, and in Section~\ref{sec:summary} we summarize.

\section{Model System}
\label{sec:model}

\subsection{Description and Basic Equations}

We will treat a section of a galactic disk or a young star cluster as an idealized model system, which allows us to isolate the physics of the radiation-matter interaction without worry about the complexity of a real disk or cluster. We consider a slab of gas with total surface density $\Sigma$ filling the domain $z>0$. It is subject to a constant gravitational force per unit mass $-g\hat{z}$ that is independent of the position $\vecx$. A vertical radiation flux $F = F_0 \hat{z}$ enters the domain of interest at $z=0$. We neglect the self-gravity of the gas and assume that all radiation is injected at $z=0$ (i.e.\ there are no internal sources of radiation at $z>0$ except the thermal emission of the gas itself.) 

For simplicity, we adopt the two-temperature flux-limited diffusion approximation. This allows us to interpolate between the optically thin and optically thick limits approximately without the need to track the radiation spectrum at each point, while not forcing the gas and radiation temperatures to be reach equality instantaneously in low optical depth regions where the matter and radiation are weakly coupled. In this approximation, the radiation flux $\mathbf{F}$ and energy density $E$ are related by
\begin{equation}
\label{eq:fld}
\mathbf{F} = -\frac{c \lambda}{\kappa_R \rho} \nabla E,
\end{equation}
where $\kappa_R$ is the Rosseland mean opacity (which can in general be a function of the gas temperature, the radiation energy density, and the gas density), $\rho$ is the gas density, and $\lambda$ is the dimensionless flux limiter, given in detail below. The equations of radiation hydrodynamics applicable to this case are \citep{krumholz07b}
\begin{eqnarray}
\label{eq:continuity}
\frac{\partial}{\partial t} \rho & = & -\nabla \cdot (\rho \vecv) \\
\label{eq:momentum}
\frac{\partial}{\partial t}(\rho \vecv) & = & -\nabla \cdot (\rho\vecv\vecv) - \nabla P - \lambda \nabla E - \rho g \hat{z} \\
\frac{\partial}{\partial t}(\rho e) & = & -\nabla \cdot [(\rho e+P)\vecv] - \kappa_P \rho (4\pi B - c E) 
\nonumber \\
& & {} + \lambda \left(2\frac{\kappa_P}{\kappa_R} - 1\right) \vecv \cdot \nabla E - \frac{3-R_2}{2} \kappa_P\rho  \frac{v^2}{c} E 
\nonumber \\
\label{eq:gasenergy}
& & {} - \rho g v_z \\
\frac{\partial}{\partial t}E & = & \nabla\cdot \left(\frac{c\lambda}{\kappa_R \rho} \nabla E\right) + \kappa_P \rho (4\pi B - c E)
\nonumber \\
& & {} - \lambda \left(2\frac{\kappa_P}{\kappa_R}-1\right) \vecv\cdot\nabla E
\nonumber \\
& & {} + \frac{3-R_2}{2} \kappa_P \rho \frac{v^2}{c} E - \nabla \cdot \left(\frac{3-R_2}{2}\vecv E\right),
\label{eq:radenergy}
\end{eqnarray}
where $\vecv$ is the gas velocity, $T_g$ is the gas temperature, $P = \rho k_B T_g/\mu$ is the gas pressure, $e = (\gamma-1)^{-1} k_B T_g/\mu + v^2/2$ is the gas specific energy, $\mu$ is the mean mass per gas particle, $\gamma$ is the gas ratio of specific heats, $B=c a T_g^4/4\pi$ is the frequency-integrated Planck function, $\kappa_P$ is the Planck mean opacity, and $R_2$ is the Eddington factor. For convenience we also define the radiation temperature by $T_r = (E/a)^{1/4}$. The above equations are written in the mixed frame, so $\kappa_P$ and $\kappa_R$ are to be evaluated in the frame comoving with the gas but all other quantities are evaluated in the lab frame, so that total energy is conserved. The functions $\lambda$ and $R_2$ have the property that $\lambda \rightarrow 1/3$ and $R_2\rightarrow 1/3$ in the optically thick limit, and $\lambda\rightarrow \kappa_R \rho E/\nabla E$ and $R_2 \rightarrow 1$ in the optically thin limit. We adopt the flux limiter and Eddington factor of \citet{levermore81a} and \citet{levermore84a},
\begin{eqnarray}
\lambda & = & \frac{1}{R} \left(\mbox{coth}\, R - \frac{1}{R}\right) \\
R & = & \frac{|\nabla E|}{\kappa_R \rho E} \\
R_2 & = & \lambda + \lambda^2 R.
\end{eqnarray}

\subsection{Non-Dimensionalization}
 
In order to characterize the problem better, and understand what dimensionless numbers control it, we now non-dimensionalize equations (\ref{eq:continuity}) -- (\ref{eq:radenergy}). We first note that, in any steady state configuration, the energy entering the slab at $z=0$ must match the energy escaping to $z=\infty$. As $z\rightarrow\infty$ the density must approach 0 for any physically reasonable configuration, so for some sufficiently large $z$ the gas must be optically thin, and thus as $z\rightarrow\infty$ the flux and radiation energy density must approach the relationship $\mathbf{F}_\infty = c E_\infty \hat{z}$. This motivates us to non-dimensionalize the equations by defining a reference temperature
\begin{equation}
\ts = \left(\frac{F_0}{ca}\right)^{1/4}.
\end{equation}
In steady state, both $T_r$ and $T_g$ must approach $\ts$ as $z\rightarrow \infty$.

From this temperature we also define
\begin{equation}
\cs = \sqrt{\frac{k_B \ts}{\mu \mh}}
\qquad
\hs=\frac{\cs^2}{g}
\qquad
\rho_* = \frac{\Sigma}{h_*}
\end{equation}
as the associated sound speed, scale height, and density; here $\mu$ is the mean mass per particle in hydrogen masses.\footnote{Note that this scale height is the value we would expect in the limit of negligible radiation forces, but even if $\hs$ does not describe the actual gas configuration, it still provides a convenient reference length.} With these definitions, we make a change of variables
\begin{equation}
\vecxi = \vecx/\hs \qquad s = t/t_* \qquad \red{t_* = \hs/\cs},
\end{equation}
and equations (\ref{eq:continuity}) -- (\ref{eq:radenergy}) become
\begin{eqnarray}
\label{eq:continuity_nondim}
\frac{\partial}{\partial s} b & = & -\nabla_\xi (\rho \vecu) \\
\frac{\partial}{\partial s} (b \vecu) & = & -\nabla_\xi \cdot (\rho \vecu\vecu) - \nabla_\xi (b\Theta_g) - \frac{\fE}{\tau_*} \lambda\nabla_\xi \Theta_r^4 
\nonumber \\
& &  {} - b \hat{\xi}_z \\
\frac{\partial}{\partial s}(bq) & = & -\nabla_\xi \left[b\left(q + \Theta_g \right) \vecu\right]
 - \frac{1}{3} \fE k_0 k_P b (\Theta_g^4 - \Theta_r^4)
\nonumber \\
& & {} + \frac{\fE}{\tau_*} \lambda \left(2 k_0 \frac{k_P}{k_R} - 1\right) \vecu \cdot\nabla_\xi \Theta_r^4 
\nonumber \\
& & {} - \frac{3- R_2}{2} \beta_s \fE k_0 k_P b u^2 \Theta_r^4 - b u_z 
\label{eq:gasenergy_nondim}
\end{eqnarray}
\begin{eqnarray}
\frac{\partial}{\partial s} \Theta_r^4 & = & \frac{1}{\beta_s\tau_*} \nabla_\xi \cdot \left(\frac{\lambda}{k_R b} \nabla_\xi \Theta_r^4\right)
+ \tau_* k_0 k_P b (\Theta_g^4-\Theta_r^4)
\nonumber \\
& & {}  - \lambda \left(2 k_0\frac{k_P}{k_R} - 1\right) \vecu \cdot\nabla_\xi \Theta_r^4 
\nonumber \\
& & {} + 
\frac{3-R_2}{2} \tau_* \beta_s k_0 b u^2 \Theta_r^4 
\nonumber \\
& & {}
- \nabla_\xi \left(\frac{3-R_2}{2} \vecu \Theta_r^4\right).
\label{eq:radenergy_nondim}
\end{eqnarray}
where $\nabla_\xi$ indicates spatial differentiation with respect to $\vecxi$ rather than $\vecx$, and we have defined the non-dimensional variables
\begin{eqnarray}
b =\frac{\rho}{\rhos} = \rho \frac{\cs^2}{\Sigma g} & \qquad &
\vecu = \frac{\vecv}{\cs} \\
\Theta_g = \frac{T_g}{\ts} & &
\Theta_r =  \frac{T_r}{\ts} \\
k_P = \frac{\kappa_P}{\kappa_P(\rhos,\ts)} & &
k_R = \frac{\kappa_R}{\kappa_R(\rhos,\ts)} \\
q = \frac{e}{\cs^2} = \frac{\Theta_g}{\gamma-1} + \frac{u^2}{2}.
\end{eqnarray}
These are, respectively, the dimensionless density, velocity, gas temperature, radiation temperature, Planck mean opacity, Rosseland mean opacity, and gas specific energy. The dimensionless ratio $R$ that determines the flux-limiter is given by
\begin{equation}
\label{eq:fluxlim_nondim}
R = \frac{4}{\tau_*} \frac{\left|\nabla_\xi \Theta_r\right|}{b k_R \Theta_r}.
\end{equation}
The dimensionless ratios appearing in equations (\ref{eq:continuity_nondim}) -- (\ref{eq:fluxlim_nondim}) are
\begin{eqnarray}
\fE & = & \frac{\kappa_{R,*} F_0}{gc}\\
\beta_s & = & \frac{\cs}{c} = \frac{1}{c} \sqrt{\frac{k_B}{\mu}} \left(\frac{g}{a \kappa_{R,*}} \fE\right)^{1/8} \\
\tau_* & = & \Sigma \kappa_{R,*} \\
k_0 & = & \frac{\kappa_{P,*}}{\kappa_{R,*}},
\end{eqnarray}
where $\kappa_{R,*} = \kappa_R(\rhos,\ts)$ and similarly for $\kappa_{P,*}$. These quantities have simple physical interpretations: for matter and radiation at the reference density and temperature $\rho_*$ and $T_*$, $\fE$ is the ratio of the radiative and gravitational forces, $\beta_s$ is the ratio of the sound speed to the speed of light, $\tau_*$ is the Rosseland mean optical depth of the slab of gas, and $k_0$ is the ratio of the Planck and Rosseland mean opacities. Finally, the condition that the flux at $z=0$ be $F_0$ combined with the flux-limited diffusion approximation (equation \ref{eq:fld}) requires that
\begin{equation}
\frac{d\Theta_r}{d\xi_z} = -\frac{\tau_* k_R b}{4 \lambda \Theta_r^3}
\end{equation}
at $z=0$.

An additional simplification is possible if we limit ourselves to the static diffusion or streaming limits, meaning that we require that $\beta_s \tau_* \ll 1$.\footnote{Formally, the product $\beta_s \tau_*$ determines whether the system is described by static or dynamic diffusion. For more discussion, see \citet{krumholz07b}.}  This is likely to hold in any real galactic disk or star cluster, since $\tau_*$ is never more than a few tens, while $\beta_s$ is generally of order $10^{-5}$. In this case we can drop the terms proportional to $\beta_s\tau_*$ in equations (\ref{eq:gasenergy_nondim}) and (\ref{eq:radenergy_nondim}), and these equations simplify to
\begin{eqnarray}
\frac{\partial}{\partial s}(bq) & = & -\nabla_\xi \left[b\left(q + \Theta_g \right) \vecu\right]
 - \frac{1}{3} \fE k_0 k_P b (\Theta_g^4 - \Theta_r^4)
\nonumber \\
& & {} + \frac{\fE}{\tau_*} \lambda \left(2 k_0 \frac{k_P}{k_R} - 1\right) \vecu \cdot\nabla_\xi \Theta_r^4 
- b u_z 
\label{eq:gasenergy_nondim1}
\\
\frac{\partial}{\partial s} \Theta_r^4 & = & \frac{1}{\beta_s\tau_*} \nabla_\xi \cdot \left(\frac{\lambda}{k_R b} \nabla_\xi \Theta_r^4\right)
+ \tau_* k_0 k_P b (\Theta_g^4-\Theta_r^4)
\nonumber \\
& & {}  - \lambda \left(2 k_0\frac{k_P}{k_R} - 1\right) \vecu \cdot\nabla_\xi \Theta_r^4 
\nonumber \\
& & {}
- \nabla_\xi \left(\frac{3-R_2}{2} \vecu \Theta_r^4\right).
\label{eq:radenergy_nondim1}
\end{eqnarray}

The above analysis demonstrates that for a given functional form of the dimensionless opacities $k_P$ and $k_R$, the behavior of the fluid in this simplified set up is dictated by only three dimensionless numbers\footnote{The opacity ratio $k_0$ also enters, but it is always of order unity for physically reasonable continuum opacity sources.}: the optical depth $\tau_*$, the Eddington ratio $\fE$, and the dimensionless ratio $\beta_s$. The first of these determines how effectively radiation is trapped in the slab, the second determines the dynamical importance of radiation pressure relative to gravity, and the third determines the relative importance of radiation advection to radiation emission and absorption in matter radiation coupling, though its precise value is unlikely to matter as long as $\beta_s \ll 1$. Thus in practice $\tau_*$ and $\fE$ determine the parameter space of interest. Given values for the dimensionless parameters, the dimensionless solution can then be scaled to physical units by a choice of the mean particle mass $\mu$ and the two dimensional parameters $\Sigma$ and $g$.

\subsection{Equilibrium Solutions}
\label{sec:equilibrium}

To understand the behavior of the equations, it is helpful to first search for equilibrium solutions, in which all time derivatives vanish and $\vecu = 0$. Examination of equation (\ref{eq:gasenergy_nondim}) immediately indicates that such solutions must obey $\Theta_r = \Theta_g = \Theta$. Inserting this condition and $\vecu = 0$ into the remaining equations reduces the problem to a pair of coupled, nonlinear ordinary differential equations
\begin{eqnarray}
\label{eq:equil1}
\frac{d}{d\xi_z}(b\Theta) + 4 \frac{\fE}{\tau_*} \lambda \Theta^3 \frac{d\Theta}{d\xi_z} + b & = & 0 \\
\label{eq:equil2}
\frac{d}{d\xi_z} \left(\frac{\lambda \Theta^3}{k_R b} \frac{d\Theta}{d\xi_z}\right) & = & 0.
\end{eqnarray}
Note that this system of equations depends only on $\fE$ and $\tau_*$, not on $\beta_s$ or $k_0$. The latter two quantities are relevant only when the radiation field is out of equilibrium. These equations are to be integrated from $z=0$ to $\infty$, subject to the boundary conditions
\begin{eqnarray}
\label{eq:f0}
\left.\frac{d\Theta}{d\xi_z} \right|_{z=0} & = & -\left.\frac{\tau_* k_R b}{4 \lambda \Theta^3}\right|_{z=0} \\
\lim_{z\rightarrow\infty} \Theta & = & 1,
\label{eq:thetainf}
\end{eqnarray}
which are equivalent to requiring that the flux be $F_0$ at $z=0$ and $z=\infty$. The third boundary condition required to specify this third-order system comes from the integral constraint that
\begin{equation}
\label{eq:bint}
\int_0^\infty b \, d\xi_z = 1,
\end{equation}
which is equivalent to demanding that $\int \rho\,dz = \Sigma$.

We can analytically integrate equation (\ref{eq:equil2}) once, and doing so and using the boundary condition (\ref{eq:f0}) allows us to rewrite the system in the somewhat more transparent form
\begin{eqnarray}
\label{eq:equil1a}
\frac{d}{d\xi_z}(b\Theta) & = & -\left(1 - \fE k_R\right) b \\
\label{eq:equil2a}
\frac{d\Theta}{d\xi_z} &= &-\frac{\tau_* k_R b}{4\lambda \Theta^3}.
\end{eqnarray}
These equations have a simple physical interpretation. The quantity $b\Theta$ is the dimensionless gas pressure, and $\fE k_R$ is the dimensionless version of the Eddington ratio $\kappa_R F/gc$ at a given point in the disk. Since the flux is invariant with $z$, this in turn is just the Eddington ratio at infinity, $\fE$, scaled by $k_R$, the ratio of the local opacity to the opacity at infinity. Thus equation (\ref{eq:equil1a}) simply asserts that the gas pressure gradient balances the force of gravity, diluted by radiation pressure, at every point. Equation (\ref{eq:equil2a}) asserts that the temperature gradient is such that the radiation flux is constant with height.

For a given value of $\fE$ and $\tau_*$, and a specified functional form of $k_R$, we can solve equations (\ref{eq:equil1a}) and (\ref{eq:equil2a}) via a double iteration procedure. We begin by guessing values for $b(0)$ (which must be $<1$) and $\Theta(0)$ (which must be $>1$) at $z=0$, and using these boundary conditions to integrate the system of ODEs from $z=0$ up to a value of $z$ large enough so that $d\Theta/d\xi_z \approx 0$, or until we encounter a singular point where $\Theta\rightarrow 0$ at finite $z$. In general the resulting solution will not satisfy the constraint that $\Theta\rightarrow 1$ at large $z$ (equation \ref{eq:thetainf}). We therefore hold $b(0)$ fixed and iterate on $\Theta(0)$ until we find the value of $\Theta(0)$ that does give $\Theta\rightarrow 1$ at large $z$. However, this will still not generally satisfy the integral constraint on $b$ (equation \ref{eq:bint}), and we must therefore iterate again to find a value of $b(0)$ such that equation (\ref{eq:bint}) is satisfied. For each value of $b(0)$ we must again iterate on $\Theta(0)$, giving rise to a double iteration.\footnote{In principal we could eliminate one iteration by adopting $\Theta = 1$ at some large $z$ as a boundary condition and integrating in the $-z$ direction. In this direction, however, the system of ODEs is stiff, and thus it is more computationally efficient to integrate in the $+z$ direction and iterate on both $b(0)$ and $\Theta(0)$.} 

\begin{figure}
\plotone{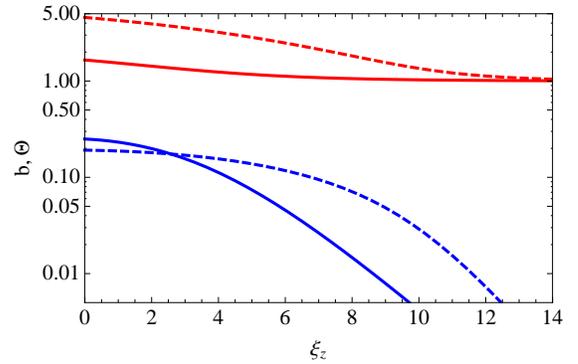}
\caption{
\label{fig:eqsolutions}
Equilibrium values for $b$ (blue) and $\Theta$ (red) as a function of $\xi_z$, computed for an opacity law $k_R = \Theta^2$, with $\fE = 0.3$ and $\tau_* = 1$ (solid) and $\fE = 0.03$ and $\tau_* = 10$ (dashed). The Eddington ratios at $\xi_z = 0$ for these solutions are 0.82 and 0.49, respectively.
}
\end{figure}

Figure \ref{fig:eqsolutions} shows two sample solutions, both computed with $k_R = \Theta^2$, the functional form that approximately describes dust opacity in the infrared \citep{semenov03a}. The two solutions shown correspond to $\fE = 0.3$, $\tau_* = 1$ and $\fE = 0.03$, $\tau_* = 10$. These choices do not map to a unique set of physical parameters, but as an example choice we can consider the same flux value and opacity law that we will use in our numerical experiments below. The flux is $F_0 = 2.6\times 10^{13}$ $L_\odot$ kpc$^{-2}$, appropriate for a bright ULIRG; the corresponding value of $T_* = 82$ K, and this gives $\kappa_{R,*} = 2.1$ cm$^2$ g$^{-1}$. With this opacity and radiation flux, the remaining dimensional parameters are $\Sigma = (0.47, 4.7)$ g cm$^{-2}$, $g = (2.5\times 10^{-6}, 2.5\times 10^{-5})$ dyne g$^{-1}$, $h_* = (3.8\times 10^{-4}, 3.8\times 10^{-5})$ pc, and $\rho_* = (4.0\times 10^{-16}, 4.0\times 10^{-14}$ g cm$^{-3})$, where the first numerical value given in parentheses is for $\tau_* = 1$, $\fE=0.3$, and the second is for $\tau_*=10$, $\fE=0.03$.

The qualitative behavior of the solutions is straightforward to understand. At large $\xi_z$ where the gas is optically thin, the equilibrium dimensionless temperature approaches a constant value $\Theta = 1$. However,  the temperature at the midplane ($\xi_z = 0$) is higher because the gas is optically thick. The increase in temperature at the midplane is largest for the case $\tau_* = 10$, since a higher optical depth leads to more effective trapping of the radiation. In both cases, the increased temperature leads to a decreased density $b$ at $\xi_z = 0$ compared to $b = 1$, the value that would be produced if the gas were isothermal. In addition, the density declines much more slowly with $\xi_z$ than would be the case for a simple isothermal atmosphere; the scale height is $\sim 5-10$, compared to $1$ for a simple isothermal atmosphere. The more gradual falloff in the density arises from two effects. First, since the Eddington ratio near the midplane is a significant fraction of unity, radiation pressure force helps support the atmosphere against gravity. Second, the gas near the midplane is hotter than for a simple isothermal atmosphere, further increasing the scale height. Both of these effects are larger in the $\fE = 0.03$, $\tau_* = 10$ case due to the greater optical depth of the atmosphere, even though the radiation pressure force at the top of the atmosphere in this case is smaller than in the $\fE = 0.3$, $\tau_* = 1$ case. Both of these effects are strongest at small values of $\xi_z$ where the temperature, radiation energy density, and radiation force are elevated. At larger $\xi_z$, where $\Theta\approx 1$, radiation pressure force becomes weak, the gas is close to isothermal, and the density falls off rapidly, returning to the behavior expected for a normal isothermal atmosphere.

\begin{figure}
\plotone{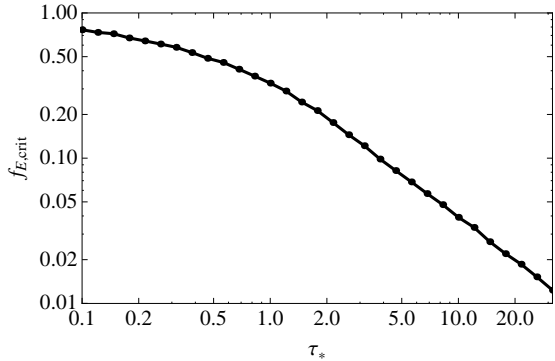}
\caption{
\label{fig:fEmax}
$\fEc$, the maximum value of $\fE$ at the slab surface for which equilibrium is possible, as a function of slab optical depth $\tau_*$, using an opacity law $k_R = \Theta^2$. Circles indicate values of $\tau_*$ for which we numerically determined values of $\fEc$.
}
\end{figure}

The iteration procedure we use to generate these solutions is not guaranteed to converge, because an equilibrium profile for which the density and temperature remain finite as $z\rightarrow \infty$ is not guaranteed to exist for an arbitrary combination of $\fE$, $\tau_*$, and $k_R$. In particular, note that if $\fE \geq 1$, then since $k_R \rightarrow 1$ as $\xi_z \rightarrow \infty$, it follows that the right hand side of equation (\ref{eq:equil1a}) is non-negative as $\xi_z\rightarrow \infty$. In this case $\int b \, d\xi_z$ diverges, and there is no solution possible that satisfies condition (\ref{eq:bint}). Moreover, even if $\fE < 1$, there may still be no solution that obeys both the constraint equations (\ref{eq:thetainf}) and (\ref{eq:bint}). In practice the maximum  value of $\fE$ for which an equilibrium solution exists, which we refer to as $\fEc$, must be determined numerically for a given functional form of $k_R$ and optical depth $\tau_*$. Figure \ref{fig:fEmax} shows the results of such a calculation for $k_R = \Theta^2$.

\section{Numerical Simulations}
\label{sec:simulations}

\begin{deluxetable*}{cccccccccc}
\tabletypesize{\scriptsize}
\tablecaption{Simulation Physical Parameters \label{tab:simulations}}
\tablehead{
& \colhead{\qquad\qquad}
& \multicolumn{3}{c}{Dimensionless Parameters}
& \colhead{\qquad\qquad}
& \multicolumn{4}{c}{Dimensional Parameters}
  \\
\colhead{Run Name} &
\colhead{} &
\colhead{$\tau_*$} &
\colhead{$\fE$} &
\colhead{$\fE/\fEc$} &
\colhead{} &
\colhead{$\Sigma$} &
\colhead{$g$} &
\colhead{$h_*$} &
\colhead{$t_*$} \\
\colhead{} &
\colhead{} &
\colhead{} &
\colhead{} &
\colhead{} &
\colhead{} &
\colhead{[g cm$^{-2}$]} &
\colhead{[$10^{-6}$ dyne g$^{-1}$]} &
\colhead{[$10^{-4}$ pc]} &
\colhead{[kyr]} \\
}

\startdata
T10F0.02 & & 10 & 0.02 & 0.51 & & 4.7 & 37 & 0.25 & 0.045 \\
T03F0.50 & & 3 & 0.5 & 3.8 & & 2.9 & 3.7 & 2.5 & 0.46 \\
T10F0.25 & & 10 & 0.25 & 6.4 & & 4.7 & 2.9 & 3.2 & 0.57 \\
T10F0.50 & & 10 & 0.5 & 12.8 & & 4.7 & 1.5 & 6.3 & 1.1 \\
\red{T10F0.90} & & \red{10} & \red{0.9} & \red{23.1} &  & \red{4.7} & \red{0.83} & \red{11} & \red{2.1} \\
\enddata
\tablecomments{All simulations use $\beta_s=1.8\times 10^{-6}$ and $k_0=10^{1/2}$, which correspond to dimensional parameters $T_* = 82$ K and $F_0 = 2.6\times 10^{13}$ $\lsun$ kpc$^{-2}$. Dimensionless and dimensional parameters are related by using the opacity law given by equation (\ref{eq:kappa}), and adopting a mean particle mass $\mu = 2.33$, as expected for a gas of molecular hydrogen and helium mixed in the standard cosmic abundance. Values of $\fEc$ have been computed numerically; for $\tau_* = 3$,  $\fEc=0.13$, and for $\tau_* = 10$, $\fEc=0.039$.
}
\end{deluxetable*}

Now that we have identified the important dimensionless numbers and found equilibria whenever they exist, we turn to a full numerical simulation that will allow us to explore the behavior of non-equilibrium systems.

\subsection{Numerical Method}

Our simulations use the \textsc{orion} radiation-hydrodynamics code \red{\citep{klein99a, fisher02a, krumholz07b}}; the properties of this code may be found elsewhere in the literature, so we simply summarize them here. \textsc{orion} solves the equations of radiation-hydrodynamics (equations \ref{eq:continuity} -- \ref{eq:radenergy}) in the two-temperature flux-limited diffusion approximation. The code uses the conservative operator-splitting scheme of \citet{krumholz07b} to separate the implicit radiation and explicit hydrodynamic updates. The latter uses a high-order Godunov method that requires very little artificial viscosity (see \citealt{klein99a} for details). The former uses the \citet{shestakov08a} pseudo-transient continuation method to solve the implicit radiation system. For the purposes of the simulations here, we do not use \textsc{orion}'s additional capabilities for self-gravity and sink particles, and instead we impose a constant gravitational acceleration in the $-z$ direction, per equation (\ref{eq:momentum}). Although \textsc{orion} has adaptive mesh refinement capability, we do not use the AMR in these simulations, because we find that in most simulations the dense gas occupies a large fraction of the simulation volume for much of the computation time. This negates the computational advantage from using AMR.

\subsection{Choice of Simulation Parameters}

In all simulations we adopt an opacity law
\begin{equation}
\label{eq:kappa}
\kappa_{(R,P)} = (10^{-3/2}, 10^{-1}) \left(\frac{T}{10\,{\rm K}}\right)^2\mbox{ cm}^2\mbox{ g}^{-1},
\end{equation}
which is roughly in accord with the model of \citet{semenov03a} at temperatures $\la 150$ K. We choose not to use the full \citeauthor{semenov03a}~opacity function (which \textsc{orion} includes) in order to keep the problem as pure and simple as possible.\footnote{For numerical reasons we cap the opacity at the value corresponding to $T=10 T_*$, and we set $\kappa_{(R,P)}=0$ in material with density $<10^{-10}\rho_*$. These choices allow us to introduce a hot, zero-opacity ambient medium that can act as a constant pressure boundary condition at high altitudes.} We note that this approximation will, if anything, lead us to overestimate the strength of matter-radiation coupling, since we are not including the flattening of the opacity at high temperatures.

For our choice of other parameters, we pick values that span an interesting physical range, and that overlap with observations. On the latter point, observations of ULIRGs and the models to fit them provided by \citet{thompson05a} give typical fluxes $10^{13}-10^{14}$ $\lsun$ kpc$^{-2}$, typical surface temperatures of $T \sim 50-100$ K, and typical Rosseland mean optical depths at this temperature are $\tau_* \sim 1-10$. The corresponding surface densities are $\sim 1-10$ g cm$^{-2}$, and the corresponding gravitational acceleration is $g = 2\pi G \Sigma \sim 10^{-6}-10^{-5}$ dyne g$^{-1}$, where we have computed $g$ as for an infinite slab, and we have not included stellar mass. Combining these estimates we find values of $\fE$ in the range $0.01 - 1$. If we focus on young clusters we find similar surface densities and fluxes, but with a somewhat larger range, so that $\fE > 1$ in some cases \citep[e.g.][]{krumholz09d}. Physically, the most interesting range to explore is the one where $\fEc < \fE < 1$. (Note that this regime exists only if $\tau_* \ga 1$, since $\fEc$ is significantly less than unity only if the gas is optically thick). These are the cases where there is no stable equilibrium, but the radiation force is not so strong that all the gas is certain to be ejected; instead the outcome will depend on how the radiation and matter couple. The majority of ULIRGs and some young, bright clusters are in this regime.

Based on these considerations, we select a set of run parameters described in Table \ref{tab:simulations}. These parameters have a range of optical depths $\tau_*$ and Eddington factors $\fE$. We include one case in the stable regime, $\fE<\fEc$, as a control, and the remaining runs are in the unstable, $\fE>\fEc$, regime. We do not include runs with $\fE > 1$. Partly this is because the outcome in this case seems almost certain to be large-scale ejection, and partly because this regime does not appear to be reached in observed ULIRGs, although the latter statement is subject to significant uncertainties on the dust opacity and the CO to H$_2$ conversion factor in ULIRGs, both of which affect estimates of $\fE$.

\subsection{Simulation Setup}

\begin{deluxetable}{ccccc}
\tablecaption{Simulation Numerical Parameters \label{tab:simnum}}
\tablehead{
\colhead{Run Name} &
\colhead{$\Delta x/h_*$} &
\colhead{$[L_x \times L_z]/h_*$} &
\colhead{$[N_x \times N_z]$} &
\colhead{$t_{\rm run}/t_*$} \\
}
\startdata
T10F0.02 & 0.5 &  $512\times 256$ & $1024\times 512$ & 78 \\
T03F0.50 & 0.5 & $512\times 1024$ & $1024\times 2048$ & 217 \\
T10F0.25 & 0.5 & $512\times 2048$ & $1024\times 4096$ & 210 \\
T10F0.50 & 0.5 & $512\times 2048$ & $1024\times 4096$ & 215 \\
\red{T10F0.90} & \red{0.5} & \red{$512\times 4096$} & \red{$1024\times 8192$} & \red{205} \\
\red{T10F0.25\_LR\tablenotemark{a}} & \red{1.0} & \red{$512\times 2048$} & \red{$512\times 2048$} & \red{276} \\
\enddata
\tablecomments{Col.~2: Grid spacing. Col.~3: Size of the computational domain in the $x$ and $z$ directions. Col.~4: Number of computational cells in the $x$ and $z$ directions. Col.~5: Time for which the simulation was run.
}
\tablenotetext{a}{\red{Physical parameters for this run are identical to those given for run T10F0.25 in Table \ref{tab:simulations}.}}
\end{deluxetable}

All our simulations are two-dimensional Cartesian, taking place in the $(x,z)$ plane. The force of gravity is in the $-z$ direction. The numerical resolution and other run parameters are described in Table \ref{tab:simnum}, \red{and we describe a resolution study we conducted to ensure that the resolution is adequate in the Appendix}. Our boundary conditions are periodic in the $x$ direction. At $z=0$, we use reflecting boundary conditions on the hydrodynamic variables, and Neumann boundary conditions on the radiation field, with the radiation flux into the box set to the value corresponding to the parameters indicated in Table \ref{tab:simulations}. At the upper $z$ boundary, for our hydrodynamic boundary condition we place immediately outside the computational domain a low density ambient medium with density $10^{-13} \rho_*$, velocity 0, and temperature $10^3 T_*$; this places it in pressure balance in the initial condition (see below), but allows dense matter with an upward velocity to leave the computational domain freely. For the radiation, we impose a Dirichlet boundary condition that the radiation energy density is $E = a T_*^4$.

We initialize all runs with a gas density distribution
\begin{equation}
\rho = 
[1+A_p \sin(2\pi x/\lambda_p)]\left\{
\begin{array}{ll}
\rho_* e^{-z/h_*}, & e^{-z/h_*} > 10^{-10} \\
10^{-10} \rho_*, & e^{-z/h_*} \le 10^{-10}
\end{array}
\right.,
\end{equation}
a temperature $T = T_*$, and a velocity of 0, where $A_p$ and $\lambda_p$ are the amplitude and wavelength of a perturbation we introduce to seed instability. In all simulations we use $A_p=0.25$ and $\lambda_p=256h_*$. We initialize the radiation energy density to $E=a T_*^4$ at all points. Thus in the absence of a radiation flux entering the domain at $z=0$, and for $A_p=0$, the region with density $> 10^{-10}\rho_*$ would simply be a stable, isothermal atmosphere. Setting $A_p$ to a non-zero value but leaving the temperature and radiation field fixed would result in a series of stable gravity waves propagating through the atmosphere.

\section{Simulation Results}
\label{sec:simresults}

\subsection{The Stable Case}

\begin{figure}
\plotone{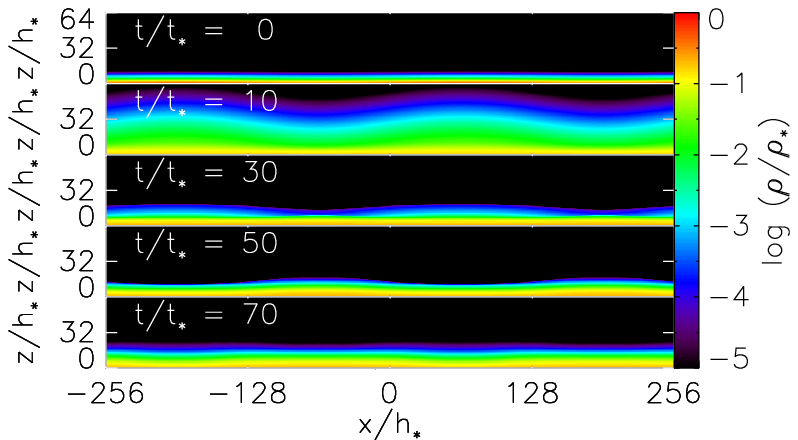}\\
\plotone{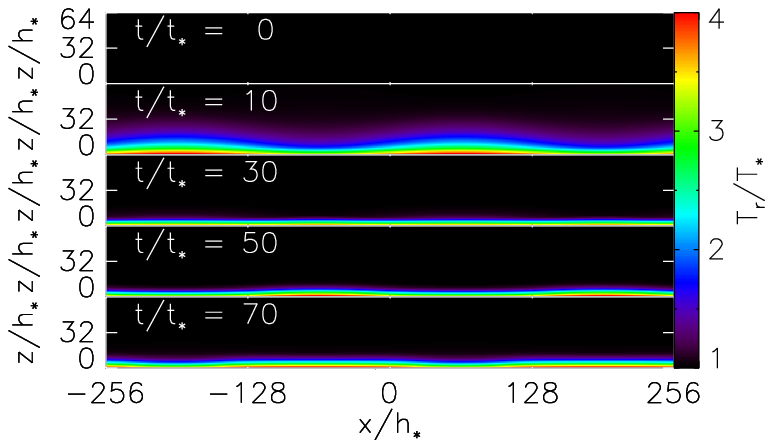}
\caption{
\label{fig:stableimg}
Time series showing the density (top) and radiation temperature (bottom) as a function of time in run T10F0.02. Gas temperatures are nearly identical to radiation temperatures everywhere except where $\rho/\rho_* < 10^{-10}$, where the set the opacity to zero. Note that the simulation domain extends to $256h_*$ in the vertical direction, but we show only the region from 0 to $64h_*$.
}
\end{figure}

\begin{figure}
\plotone{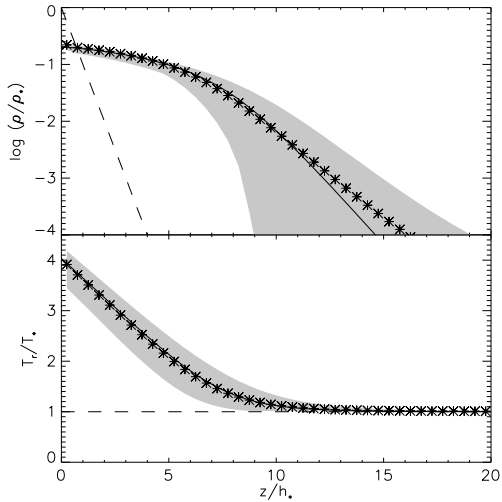}
\caption{
\label{fig:stablecomp}
Comparison between run T10F0.02 and the analytic solution for $\tau_*=10$, $\fE=0.02$, computed via the procedure described in Section \ref{sec:equilibrium}. The top panel shows density and the bottom shows radiation temperature; gas temperature is nearly identical. In both panels, the solid line is the analytic equilibrium solution, and asterisks show the mean density/temperature at a given vertical position in the simulation, with the mean computed over all horizontal positions and all simulations times $t/t_*>70$. The gray band shows the range of values found at a given vertical position over all times and horizontal positions. The dashed lines show the initial conditions in the simulation.
}
\end{figure}

We first examine run T10F0.02 \red{(recall our naming convention that T10 means $\tau_* = 10$, and F0.02 means $\fE = 0.02$)}, which we \red{may} think of as a sort of control, in the sense that $\fE/\fEc < 1$. This simulation therefore has a stable configuration toward which it is able to converge. Figure \ref{fig:stableimg} shows a series of snapshots of the density and temperature distribution in the run. As the plot shows, the gas is initially pushed upward by radiation pressure. This is not surprising, because the initial configuration is an equilibrium appropriate to the case of an isothermal gas with no radiation pressure. Once radiation is injected into the simulation domain, the temperature near $z=0$ reaches $\sim 4-5 T_*$. Since the opacity varies as $\kappa_R \propto T^2$, this means that the opacity and the radiative force felt by this gas is increased by a factor of $\sim 20$ compared to gas at a temperature $T_*$, and the local Eddington ratio in this gas approaches $\sim 0.5$. This accounts for the upward movement. Since radiation is trapped more effectively at horizontal locations where the initial density is somewhat higher, due to the initial perturbation, the upward motion is strongest there. This amplifies the initial perturbation, so that the horizontal variation visible at $t/t_* = 10$ and 20 is much larger than that present in the initial conditions. After the initial transient however, the gas settles back down, and the horizontal fluctuations damp out. There is little change past $t/t_* \sim 50$.

In Figure \ref{fig:stablecomp} we compare the state of the system at late times to the equilibrium configuration for $\tau_*=10$ and $\fE=0.02$. As the plot shows, the density and temperature converge to the equilibrium solution very well. There is some oscillation at high altitudes as the system rings down, but the convergence is clear. It is worth noting that, even though $\fE/\fEc < 1$ and $\fE = 0.02$, radiation pressure is non-negligible in the equilibrium configuration. Since $\kappa_R\propto T^2$, the peak temperature of $T \approx 4 T_*$ at the base of the simulation domain corresponds to a local Eddington ratio $f_{\rm E}\approx 1/3$.

\subsection{Unstable Cases: Morphology and Overall Evolution}
\label{sec:morphology}

\begin{figure*}
\epsscale{0.7}
\plotone{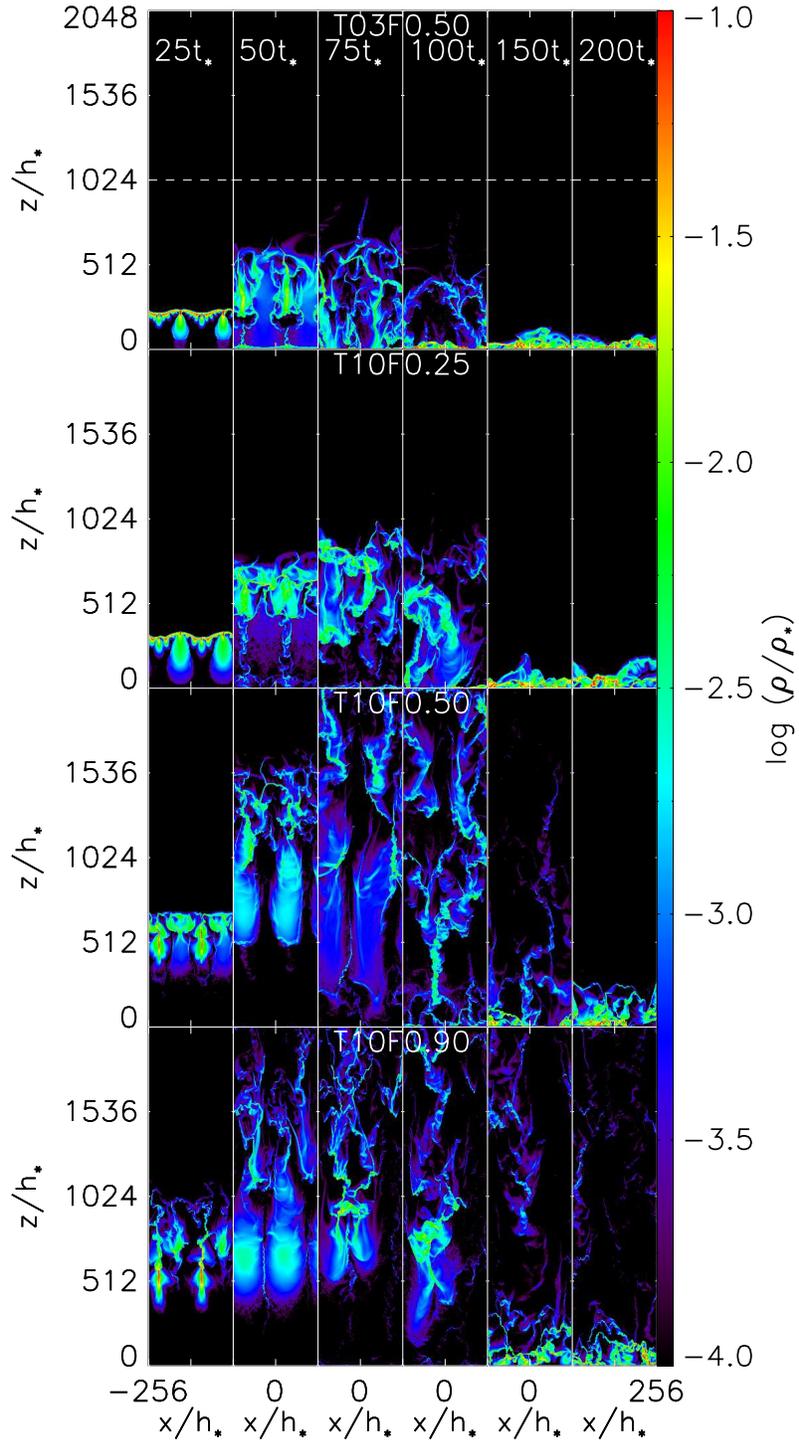}
\epsscale{1.0}
\caption{
\label{fig:rhoplot}
Gas density distribution as a function of time in simulations T03F0.25, T10F0.25, T10F0.50, \red{and T10F0.90}, as indicated. Each row represents a simulation, and each column is at a fixed time, as indicated. Blank panels indicate that the simulation was halted before the indicated time. \red{For run T03F0.25, the top of the computational domain only extends to $1024h_*$, indicated by the dashed line. For run T10F0.90, the computational domain extends to $4096h_*$, i.e.~twice the vertical extent shown.} Note that the dynamic range of densities in the simulation is much larger than the range $10^{-4}-10^{-1} \rho_*$ that we show. We pick this range because $\sim 95\%$ of the mass in the simulation volume lies within it at most times.
}
\end{figure*}

\begin{figure*}
\epsscale{0.7}
\plotone{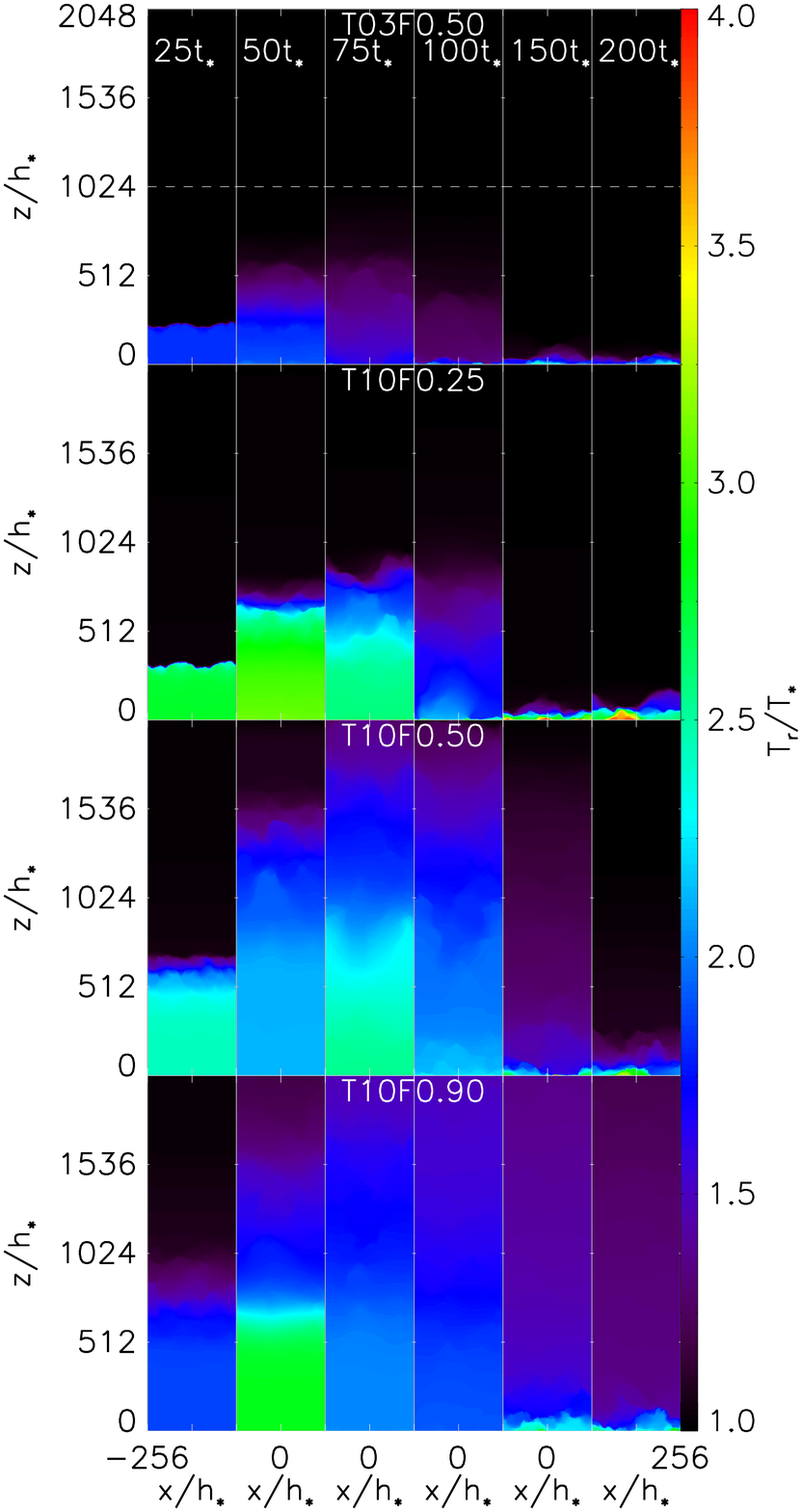}
\epsscale{1.0}
\caption{
\label{fig:tplot}
Same as Figure \ref{fig:rhoplot}, but showing the radiation temperature $\Theta_r = T_r / T_*$. The gas temperature distribution is very similar everywhere except at very high altitudes and low densities.
}
\end{figure*}

We now turn to runs T10F0.25, T03F0.50, T10F0.50, \red{and T10F0.90}, for which no equilibrium solution exists. Figures \ref{fig:rhoplot} and \ref{fig:tplot} show the time evolution of the gas and radiation energy density distributions in each of them. As in run T10F0.02, the gas is initially flung upward by radiation pressure. The gas is driven into a thin shell, with high radiation temperature beneath it and low radiation temperature above it. As time passes, however, the shell begins to buckle, developing fingers of gas that penetrate down into the radiation-dominated lower region. The buckling is evident by $t/t_*\sim 5$, fingers appear by $t/t_*\sim 10-20$, and by $t/t_*\sim 50$ a clear non-linear instability has set in. The dense fingers of gas reaching into the radiation-dominated region have much smaller upward velocities than the rest of the gas, and at late times they begin to fall back toward $z=0$.

As the instability develops the gas becomes turbulent. It also develops a clear channel morphology, with most of the gas mass in dense, nearly vertical filaments, and most of the volume filled by low-density gas between the filaments. Gas in the dense filaments is predominantly falling rather than rising. As a result, in runs T03F0.50 and T10F0.25 none of the mass reaches the upper boundary of the computational domain. In runs T10F0.50 \red{and T10F0.90} some mass does reach the top of the computational domain, but it is a small fraction of the total, and most of it is simply coasting from the initial launch at early times, rather than being actively driven upward. By $t/t_* \sim 100-150$, the upward motion has completely halted and the gas has fallen back to the midplane. Thereafter the system exhibits continuous turbulent motion with a roughly constant velocity dispersion and gas scale height. This appears to be the final, statistical steady state.

\subsection{Turbulence}

\begin{deluxetable*}{cccccccc}
\tablecaption{Simulation Outcomes \label{tab:simresult}}
\tablehead{
\colhead{Run Name} &
\colhead{$ \sigma_x/\cs$} &
\colhead{$ \sigma_z/\cs$} &
\colhead{$ \sigma/\cs$} &
\colhead{$\fEavg$} &
\colhead{$\ft$} &
\colhead{$f_{\rm trap,w}$} &
\colhead{$\kappa(T_{\rm mp})\Sigma\approx \tau_{\rm IR}$} \\
}
\startdata
T10F0.02 & 0.22 & 0.13 & 0.27 & 0.18 & 88 & 35 & 160\\
T03F0.50 & 3.2 & 2.6 & 4.1 & 1.0 & 5.0 & 2.5 & 15\\
T10F0.25 & 4.8 & 3.3 & 5.8 & 1.0 & 39 & 25 & 93\\
T10F0.50 & 6.5 & 5.9 & 8.8 & 1.1 & 22 & 13 & 120\\ 
\red{T10F0.90} & \red{5.7} & \red{17.1} & \red{18.1} & \red{1.2} & \red{12} & \red{7.0} & \red{64}  \\
\red{T10F0.25\_LR}\tablenotemark{a} & \red{4.2} & \red{2.9} & \red{5.1} & \red{1.0} & \red{40} & \red{24} & \red{110}
\enddata
\tablecomments{All quantities shown represent time averages over all times $t/t_* > 175$ in all runs except T10F0.02, where the average is over $t/t_*>50$. The quantity $\ft$ is the trapping factor considering all material (as defined by equation \ref{eq:ftrap}), while $f_{\rm trap,w}$ is the trapping factor computed considering only material with $v_z>0$.  The quantity $\kappa(T_{\rm mp})\Sigma\approx \tau_{\rm IR}$ is the average 
optical depth at the end of the calculation, computed using the mass-weighted mean midplane temperature. (Volume-weighting gives a nearly identical result.) 
Values of $\tau_{\rm IR}$ at the start of the calculation are very similar.
}
\tablenotetext{a}{\red{Physical parameters for this run are identical to those given for run T10F0.25 in Table \ref{tab:simulations}.}}
\end{deluxetable*}

\begin{figure}
\plotone{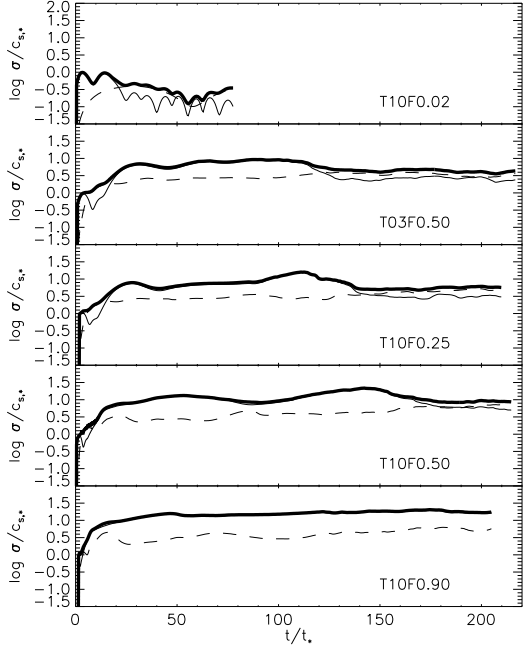}
\caption{
\label{fig:vdisp}
Velocity dispersion $\sigma$ versus time in runs T10F0.02, T03F0.50, T10F0.25, T10F0.50, \red{and T10F0.90}, as indicated in each panel. We show the $x$ and $z$ components $\sigma_x$ and $\sigma_z$ (thin dashed and solid lines, respectively), and the total velocity dispersion $\sigma=(\sigma_x^2+\sigma_z^2)^{1/2}$ (thick solid line).
}
\end{figure}

One of the important questions regarding radiation instabilities is whether they can explain, or at least contribute to, the large turbulent velocity dispersions seen in radiation-dominated galactic disks. To address this question we compute the mass-weighted horizontal and vertical velocity dispersions $\sigma_x$ and $\sigma_z$ as a function of time in our simulations. We define these by
\begin{eqnarray}
\sigma_{(x,z)} & = & \left[\frac{1}{M}\int \rho (v_{(x,z)}-\overline{v}_{(x,z)})^2 \, dV\right]^{1/2} \\
\overline{v}_{(x,z)} & = &  \frac{1}{M}\int \rho v_{(x,z)} \, dV
\end{eqnarray}
where $M$ is the amount of mass in the simulation domain and the integrals are over the full simulation volume.

Figure \ref{fig:vdisp} shows the result, and Table \ref{tab:simresult} summarizes what is shown in the Figure. Not surprisingly, in the stable run T10F0.02, after the initial transient the velocity dispersion is highly subsonic. It is dominated by the horizontal component, and oscillates up and down as the system rings down toward equilibrium. In the other runs we see that, after the initial transient, the velocity dispersion approaches a roughly constant, supersonic value. The vertical component of the velocity dispersion is somewhat larger than the horizontal one. The highest Mach numbers we reach are $\sim 10$.

\subsection{Eddington Ratio and Trapping Factor}

\begin{figure}
\plotone{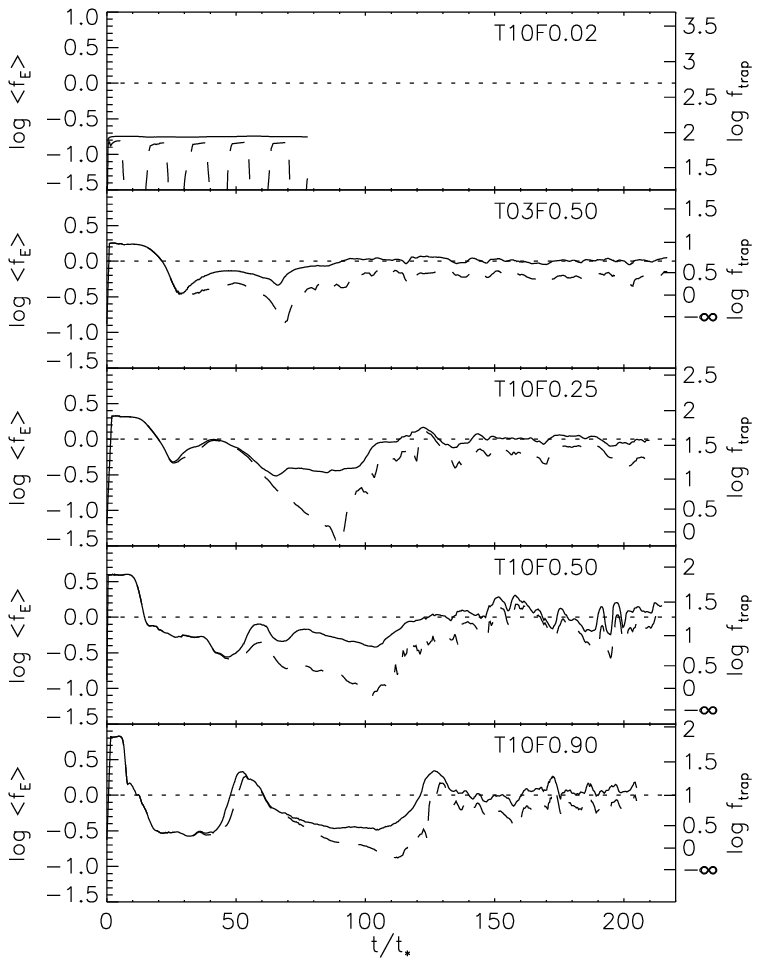}
\caption{
\label{fig:frad}
Mean Eddington ratio $\fEavg$ (left axis) and trapping factor $\ft$ (right axis) as a function of time for runs T10F0.02, T03F0.50, T10F0.25, T10F0.50, \red{and T10F0.90}, as indicated in each panel. The solid line is the radiation force applied to all the matter in the computational domain, and the dashed line is the force applied only to matter with $v_z>0$. The dotted horizontal line marks $\fEavg = 1$. Note that all simulations begin with $\fEavg = 0$ because we initialize the simulation with no gradient in the radiation energy density; the rapid rise of $\fEavg$ to its initial plateau is the result of the radiation field reaching equilibrium rapidly, on a timescale that is not resolved in the plot.
}
\end{figure}

Another quantity of interest is the net force applied by the radiation to the gas, and how this compares to the force of gravity. A closely related question is how much momentum the gas is able to extract from the radiation field and transfer into a wind. Recall that the radiation force per unit volume exerted on the matter is
\begin{equation}
\mathbf{f}_{\rm rad} = \frac{\kappa_R \rho \mathbf{F}}{c} = -\lambda \nabla E,
\end{equation}
where $\mathbf{F}$ is the radiation flux, and the second equality follows from the flux-limited diffusion approximation. The mean vertical radiation force per unit area applied to the gas in the simulation domain is
\begin{equation}
\fzavg = \frac{1}{L_x} \int \mathbf{f}_{\rm rad} \cdot \hat{z} \, dV,
\end{equation}
where $L_x$ is the horizontal length of the computational domain, and the integral is over the full simulation volume. It useful to compare this force to two other quantities. One is the mean force exerted by gravity, $f_g = -\Sigma g_z$. We define the mean Eddington ratio of the computational volume by
\begin{equation}
\fEavg = \frac{\fzavg}{\Sigma g_z}.
\end{equation}
The bulk of the matter can be ejected by radiation pressure only if $\fEavg > 1$, although a wind containing a small fraction of the matter may be driven even if $\fEavg < 1$.

The second useful comparison is between the force exerted by the radiation and the momentum flux carried by the radiation field, which is $F_0/c$. As discussed in Section~\ref{sec:intro}, if the medium is optically thick enough to ensure that every photon is absorbed at least once (as is the case for our simulations), we expect the radiation field to transfer at least this much momentum to the gas. However, the amount of momentum extracted could be significantly larger if every photon is absorbed many times, up to a maximum value of order $F_0/v$, where $v$ is the gas characteristic velocity. To quantify where between these two limits our simulations fall, we follow \citet{krumholz09d} and define the trapping factor $\ft$ by
\begin{equation}
\label{eq:ftrap}
\fzavg = (1+ \ft) \frac{F_0}{c} .
\end{equation}
In the limit of one absorption per photon we expect $\ft = 0$, and for infinitely many absorptions we expect $\ft \approx c/v$. Models that assume strong trapping generally adopt $\ft \sim \tau_{\rm IR}$, where $\tau_{\rm IR}$ is a fiducial optical depth that depends on the mean gas temperature. For example, in their numerical simulations \citet{hopkins11a} adopt $\ft=\max(0,\kappa_0 \Sigma-1)$, where $\kappa_0 = 5$ cm$^2$ g$^{-1}$ and $\Sigma$ is the gas surface density; values of $\kappa_0\Sigma$ in the simulations range from $\ll 1$ up to $\sim 50$. Finally, note that $\fEavg$ and $\ft$ are related by
\begin{equation}
\label{eq:feavg}
\fEavg = (1 + \ft) \frac{\fE}{\tau_*}.
\end{equation}

We show $\fEavg$ and $\ft$ for our simulations in Figure \ref{fig:frad}, and summarize the mean values in Table \ref{tab:simresult}. The are several interesting points to be taken from this plot. First, in the stable run T10F0.02, $\fEavg < 1$ at all times. This is not surprising, since the run parameters were chosen to have this feature. In contrast, in all of the unstable runs the pattern is that $\fEavg > 1$ at early times before the instability sets in. Once the instability is established, $\fEavg$ drops to less than unity, and the gas falls back. Finally, at late times when the instability is in steady state, $\fEavg$ is very close to unity. The value of $\fEavg$ in this steady state does not depend on $\fE$ or $\tau_*$. Thus we see that, even though $\tau_* \fE > 1$ in all three of the unstable runs, the actual force applied to the matter by the radiation field saturates at the Eddington limit. Thus there is no large scale ejection of matter, despite the fact that $f_{\rm Edd}$ is much larger than unity at the midplane at the start of the calculation.

A similar effect is apparent if we consider the trapping factor $\ft$. In the stable run, T10F0.02, $\ft\sim 100$. This is because the radiation field is effectively trapped by the uniform matter distribution, so every photon is absorbed and re-emitted many times. In contrast, in the unstable runs $\ft$ is significantly smaller once the instability develops. Roughly half this force is applied to material with an upward velocity, simply because roughly half the gas has a positive velocity at any given time. However, the identity of individual upward-moving and downward-moving pockets of gas is continually changing, so no material has a sustained positive velocity and is launched into a wind. A reasonable conjecture based on the simulation results is that in the regime where $\fE < 1 < \tau_* \fE$, the gas self-adjusts to ensure $\fEavg \approx 1$, and that $\ft$ goes to whatever value is required to make this happen. In this case we would predict that $\ft \approx \tau_*/\fE-1$. However, all of this momentum is delivered to bound gas, and does not produce a wind.

\subsection{\red{Conversion of Radiative Power to Kinetic, Thermal, and Gravitational Potential Energy}}

A final quantity of interest is how much radiative power is converted to the kinetic, thermal, and gravitational potential energy of the gas. At late times, once the turbulence has reached steady state, the net power transfer rate between the various energy reservoirs must be zero. This conclusion follows simply from the fact that $\sigma$, the mean gas temperature, and the gas scale height all approach statistically constant values at late times, implying that the kinetic, thermal, and gravitational potential energies of the gas must be statistically constant as well. In practice the system produces this effect via a balance between two processes. The radiation exerts forces on the gas, doing work; this effect is represented by the equal and opposite terms proportional to $\vecv$ and $E$ in Equations (\ref{eq:gasenergy}) and (\ref{eq:radenergy}). Once the gas becomes turbulent, however, regions of compression appear, and in these regions kinetic energy is converted to thermal energy, which in turn is turned back into radiative energy, a process described by the terms $\pm \kappa_P \rho (4\pi B - cE)$ in Equations (\ref{eq:gasenergy}) and (\ref{eq:radenergy}). (Both exchanges can go in the opposite direction too: gas can do work on the radiation field, and cool gas can be radiatively heated. In our problem these effects are generally smaller than their opposites, however.) These rates of exchange are on average equal and opposite, and the level of turbulence self-adjusts to ensure that they remain so.

\red{
In this respect a system that does not drive a wind, as we find in all our simulations, is fundamentally different than one that does drive a wind. In a system with a wind, the baryons that reach infinity carry kinetic, thermal, and gravitational potential energy, and thus some non-zero fraction of the radiant energy must be converted to other forms in the process of driving the wind. Without a wind, once the system reaches steady state the time-averaged net conversion of radiant energy to other forms is necessarily zero.\footnote{\red{One might object that in a planar geometry such as ours a wind can never occur, since the gravitational potential energy of a gas parcel diverges at large $\xi_z$, unlike in spherical geometry. This is true in principle but turns out to be irrelevant in practice. As we discuss in Section \ref{sec:planar}, inserting realistic astrophysical scalings into our dimensionless problem shows that the gas in our simulations reaches such small heights that curvature effects are negligible, and a planar approximation is in fact valid. Gas in our simulations is not launched into a wind not because we have placed it in a gravitational well that is infinitely deep, but because the radiation force is weak enough that it could not escape a potential well of finite, astrophysically reasonable size.}}
} This analysis leaves open, however, the related question of what level of turbulence must be created so that this steady state is achieved. Clearly the equilibrium value of $\sigma/\cs$ must be a function of $\tau_*$ and $\fE$. In this work we have sampled the $(\tau_*, \fE)$ plane only sparsely, and so we are not yet in a position to construct a model for this mapping. However, we plan to revisit this topic in future work.

\section{Discussion}
\label{sec:discussion}

\subsection{What is the Nature of the Instability?}
\label{sec:nature}

The instability that appears in our simulations is almost certainly the radiation Rayleigh-Taylor (RRT) instability, first described in simulations by \citet{krumholz09c} and formally analyzed by \citet{jacquet11a}. The instability occurs when an interface forms in a gravitational field between a low-density radiation pressure-dominated medium on the bottom and a higher density, less radiatively-dominated medium on top. We can calculate the instability's growth rate of and most unstable wavelength in the linear regime using \citeauthor{jacquet11a}'s formalism. The dispersion relation for the instability in the adiabatic, local limit (their Equation 77) in our non-dimensionalized units becomes
\begin{equation}
-\frac{C}{\Theta_g} \omega^6 + \left[\frac{C(C-B)}{\Theta_g^2} + k^2\right] \omega^4 + B \frac{k^2 \omega^2}{\Theta_g} - k^4 = 0,
\label{eq:dispersion}
\end{equation}
where
\begin{eqnarray}
x & = & \frac{a T_r^4}{3 \rho c_s^2} = \frac{\fE}{3\tau_*} \frac{\Theta_r^4}{b\Theta_g}\\
\fEl & = & \frac{\kappa_R F}{g c} = \fE k_R \left(\frac{F}{F_0}\right)\\
D & = & 16 x^2 + 20x + \frac{\gamma}{\gamma-1} \\
C & = & \frac{1}{D} \left(12 x + \frac{1}{\gamma-1}\right) \\
B & = & \frac{1}{D}\left[16 (\fEl-1)x^2 + (24 \fEl-8)x 
\right.
\nonumber \\
& & \left. \quad{}+ \fEl \left(5+\frac{\gamma}{\gamma-1}\right) - 1 + \frac{\gamma \fEl}{4(\gamma-1) x}\right].
\end{eqnarray}
Here the wavenumber $k$ is measured in units of $h_*^{-1}$ and angular frequency $\omega$ is measured in units of $g/c_{s,*}$. The quantities $x$ and $\fEl$ are the local ratio of radiation pressure to gas pressure and the local Eddington ratio, respectively, both computed in the gas immediately above the interface. The conditions of adiabaticity and locality are satisfied only if (their Equation 82)
\begin{eqnarray}
\sqrt{\min\left(1+x,\frac{x}{\fEl}\right)} & > & \frac{F/c_s}{a T_r^4 + \rho c_s^2/(\gamma-1) }
\nonumber \\
& = &
\frac{F/F_0}{\beta_s \Theta_g \left[\Theta_r^4 + \frac{b\Theta_g \tau_*}{(\gamma-1)\fE}\right]}.
\label{eq:damp}
\end{eqnarray}

To evaluate these quantities, we note that before the onset of instability $F$ is very close to $F_0$ everywhere in the computational domain, and the values of $b$, $\Theta_r$, and $\Theta_g$ as the base of the layer that goes unstable can be read off from the simulation results. Depending on the exact time we choose to examine, we find that in run T03F0.05, $b\approx 0.1 - 0.15$ and $\Theta_r \approx \Theta_g \approx 1.8$. The corresponding figures in runs T10F0.25 and T10F0.50 are $b\approx 0.15-0.2$, $\Theta_r \approx \Theta_g \approx 2.7$ and $b\approx 0.3$, $\Theta_r \approx \Theta_g \approx 2.7$. Unfortunately inserting these quantities into Equation (\ref{eq:damp}) indicates that the inequality is not satisfied, indicating that we cannot regard the modes as adiabatic, and that diffusion is significant. Nonetheless, we can still use the dispersion relation (\ref{eq:dispersion}) to obtain an upper limit on the growth rate. Inserting these figures into Equation (\ref{eq:dispersion}) and numerically solving for $\omega$, we find characteristic growth rates ${\rm Im}(\omega) \sim 1$, indicating that, if not damped by diffusion, the initial seed perturbation we insert should amplify on a timescale comparable to $c_{s,*}/g$. In practice we see that growth is significantly slower than this, almost certainly as a result of diffusive damping. Unfortunately \citet{jacquet11a} were unable to obtain an analytic estimate of the instability growth rate in the regime where diffusive damping is significant.

\red{
We can also ask whether the instability we observe in our simulations might correspond to any of the other types of radiation-hydrodynamic instability that have been described in the astrophysical literature. A number of authors have studied instabilities in radiation pressure-driven flows in the context of clouds near quasars \citep{mathews77a, krolik77a, mathews86a}. These are quite different than the situation we consider in that the opacity comes from resonant absorption of ionizing photons, which is therefore linked to the recombination rate and hence the density of the gas; instabilities arise due to this coupling. Clearly that is not the case for the situation we consider, since dust opacity is to good approximation density-independent. \citet{blaes03a} conduct a general analysis of local radiative instabilities in optically thick media applicable to a wide variety of environments. They find that local instability occurs only when there is a magnetic field present or when the opacity contains an explicit density dependence, similar to that which applies in the quasar case. Our simulations meet neither condition, and, indeed, the RRT instability described by \citet{jacquet11a} that appears to take place in our simulations is an interface instability rather than a local one. In the real ISM the dust opacity is, as noted before, density-independent. However, the real ISM does contain magnetic fields, and it is conceivable that adding a magnetic field to our simulations would produce an additional local instability on top of or instead of the interface one that we find. We discuss this issue further in Section \ref{sec:bfields}.
}

\red{
Perhaps the closest analog in the literature to our situation is that analyzed by \citet{shaviv01a}, who studies instabilities in atmospheres with a constant, Thomson, opacity. He finds that such instabilities can arise when the gas is near the Eddington limit, with the instability growth rate depending on the boundary conditions \citep{blaes03a}. The primary difference between our work and his is the radiation temperature-dependence of the dust opacity of the ISM. This leads to an Eddington ratio that is non-constant with height, which in turn means that our instability tends to take the form of gas at the bottom of the atmosphere where the radiation temperature and opacity are high being driven upward into a thin shell that collides with gas at larger altitudes that experiences a lower radiation temperature and opacity. It is this behavior that produces an interface and an interface instability. In the absence of radiation temperature-dependence in the opacity, radiative accelerations are height-independent, and thin shells should not form. In this context the global, non-interface instability of \citet{shaviv01a} can occur. The non-linear outcome of the two instabilities are likely to be different as well. In the case of an ISM with dust opacity, turbulence is driven by the circulation of material falling deep into the atmosphere, encountering a hot radiation field, finding itself super-Eddington, and then being blasted upward; once at height the radiation field pushing on the gas is at lower temperatures, the gas falls back, and the cycle repeats. This mechanism obviously cannot operate for a gray opacity.
}

\subsection{Why is Radiative Trapping Ineffective?}

\begin{figure}
\plotone{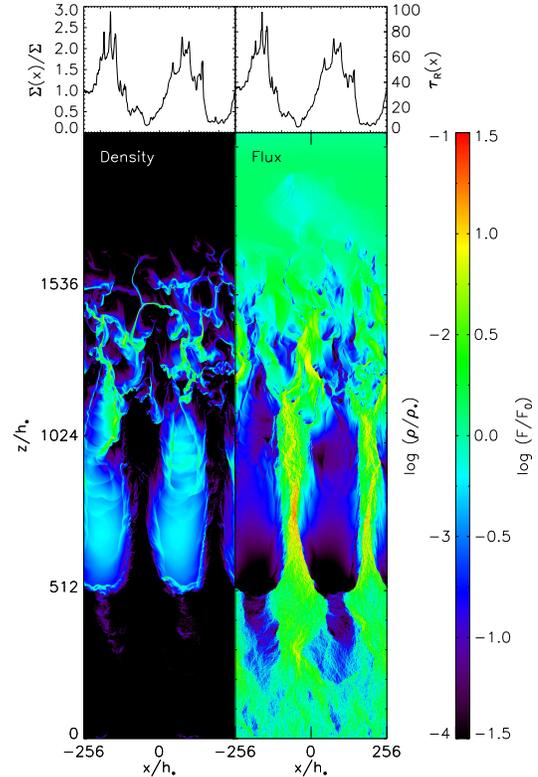}
\caption{
\label{fig:flux}
The two images show the gas density distribution (left) and magnitude of the radiation flux $F$ (right) at time $t/t_* = 50$ in run T10F0.50. The line plots above them show the surface density $\Sigma(x)$ and Rosseland mean optical depth $\tau_R(x)$ computed along vertical paths at a fixed horizontal position $x$, i.e.\ $\Sigma(x) = \int_0^\infty \rho(x,z) \, dz$ and $\tau_R(x) = \int_0^\infty \rho(x,z) \kappa_R(x,z) \, dz$.
}
\end{figure}

The relatively low values of $\fEavg$ and $\ft$ we find in our simulations are surprising. For a laminar matter distribution, even if all the gas were at a temperature $T_*$ and thus had opacity $\kappa_{R,*}$, we would have $\ft = \tau_*-1$. Considering the higher opacities produced by higher temperatures, we might expect $\ft\sim 100$, as in run T10F0.02, and $\fEavg \gg 1$. In the last column of 
Table \ref{tab:simresult} we give the value of the midplane optical depth $\tau_{\rm IR}$ calculated at the end of the simulation
using $\tau_{\rm IR} \approx \kappa(T_{\rm mp})\Sigma$. Values at the start of the calculation are very similar.
In the three unstable cases, the value of $f_{\rm trap}$ is a factor of $\sim3-6$ smaller than $\tau_{\rm IR}$, with the largest difference occurring in the most unstable run. 
How is it possible to have such a small value of $\ft$ with respect to the naive estimate? To answer this question, it is helpful to examine the distribution of radiation flux. We do so in Figure \ref{fig:flux}, using run T10F0.50 at time $t/t_*=50$ as an example. Other time slices and runs when $\ft$ is small give similar results, but this time slice, when the instability is non-linear but has not yet dissolved into complete turbulence, provides a particularly clear illustration. As the plot shows, the radiation flux is both highly non-uniform and strongly anti-correlated with the matter distribution. Within the fingers of gas projecting downward, the flux is $\sim 0.1 F_0$, while in the narrow channels between the fingers the flux is $\sim 10 F_0$. This explains how there can be so little radiation force exerted on the matter: the radiation flux is highly concentrated in low-density, low-optical depth channels that contain little mass, so the effective optical depth is much less than the true optical depth.

It is important to note, however, that this does not mean that the system is effectively optically thin, or that an external observer would be able to see directly down to the radiation source with little interfering material. As the plot shows, even along the vertical paths through the simulation domain with the lowest column densities and optical depths, corresponding to the channels through which most of the flux is focused, the Rosseland mean optical depth is always at least $\sim 5$, and the column density is always at least $\sim 0.2\Sigma$. \red{The significant point is not that the value of $\sim 5$ is particularly special; it is that the effective momentum imparted by the radiation field to the gas can be reduced compared to an estimate based on $\tau_{\rm IR}$ without there being optically thin channels that would allow direct optical observation of the stars providing the radiation flux. The absence of such transparent channels does not imply that radiation-gas coupling is efficient.}

\begin{figure}
\plotone{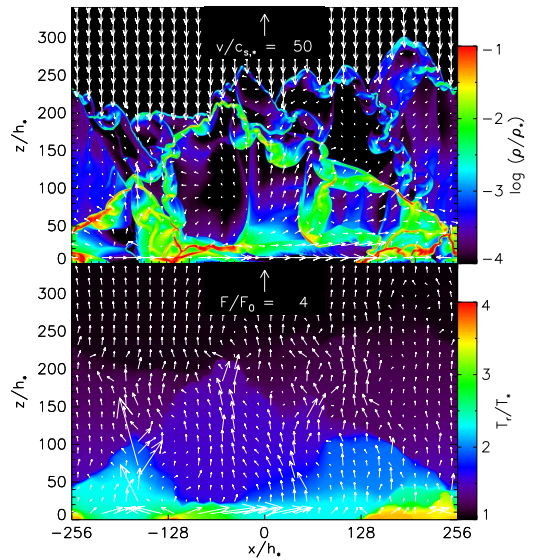}
\caption{
\label{fig:finalstate}
Density distribution and Mach number vectors (top) and temperature distribution and flux vectors (bottom) 
in the last timeslice ($t/t_*=215$) of run T10F0.50, showing a snapshot of the atmosphere's 
structure after $\langle f_{\rm E}\rangle\rightarrow1$.}
\end{figure}

While the channeling of the radiation is easiest to see early in the development of the instability, as shown in Figure \ref{fig:flux}, it continues at later times as well. Figure \ref{fig:finalstate} shows the density, temperature, velocity, and radiation flux distribution in T10F0.50 at the last time slice, $t=215t_*$. At this point the radiation pressure-driven turbulence is fully developed, and $\fEavg \approx 1$. As the plot shows, the radiation flux continues to be highly non-uniform. The structure of this atmosphere and its dependence on model parameters will be the subject of a future paper.

\subsection{The Origin of Turbulence in ULIRGs and Dense Clusters}
\label{sec:turborigin}

We find that radiation pressure-driven instabilities can produce significant turbulence, and thus the hypothesis that radiation pressure might be responsible for at least some turbulence in dense protoclusters and in the disks of ULIRGs seems to be valid. The typical Mach numbers in Galactic protoclusters are $\sim 10$ \citep[e.g.][]{shirley03a}, comparable to what we obtain in the simulations, so radiation pressure-driven instabilities may fully explain the turbulence there. (Of course we cannot rule out that other effects might contribute as well.) For ULIRGs, however, the problem is harder. Even in the most unstable run we consider, T10F0.50, the Mach number is still only $\sim 10$. This is roughly an order of magnitude less than the Mach numbers seen in real ULIRGs and those needed to maintain a Toomre $Q$ parameter near unity \citep{thompson05a}.

What then does this imply about the origin of the turbulence in ULIRGs? One possibility is that, given the observational uncertainties, ULIRG surface densities have been overestimated or ULIRG luminosities underestimated, and in fact they have $\fE \approx 1$ over large areas. In this case it seems likely that radiation pressure could be responsible for driving the turbulence, although the problem of how this process picks out $Q=1$ would still exist. A closely related possibility that answers the question of how $Q=1$ is maintained is that the turbulence is driven by a limit cycle whereby episodes of star formation produce brief periods where $\fE > 1$ and the bulk of the gas does begin to be ejected. This halts star formation, and after $\sim 4$ Myr when massive stars begin to die, $\fE$ drops below unity and the gas falls back. This ejection could either be over a large area of the disk, or it could occur locally in within forming clusters. For example, \citet{murray10a} suggest a cycle in which Toomre-mass clumps repeatedly form, undergo gravitational collapse, and then are disrupted by star clusters that rapidly reach luminosities such that $\fE > 1$. The ejecta from these disrupting clusters provide the turbulence. 

Alternately, the turbulence in ULIRGs might not be radiatively driven at all. Instead, it might be driven by gravitational instabilities in ULIRG disks \citep{krumholz10c, forbes12a}, and be maintained by the energy of inward gas mass through the galactic potential rather than by stellar feedback. The turbulence could also be driven by cosmic ray rather than radiation pressure \citep{ipavich75a, breitschwerdt91a, jubelgas08a, socrates08a}.

\subsection{Implications for Analytic and Numerical Models of Radiation Pressure Feedback}

Our work suggests that radiation pressure alone is unlikely to be able to eject much matter or drive significant winds when $\kappa(T_{\rm mp}) \rho F / c \ll \Sigma g \ll \kappa(T_{\rm edge}) \rho F / c$, i.e.~the regime where the radiation flux is sub-Eddington at the top of the disk where the temperature is low, but super-Eddington at the midplane where the temperature is high. In this regime, we find that even a very laminar initial density field will spontaneously re-arrange itself, via radiation Rayleigh-Taylor instability, into a configuration whereby the mean Eddington ratio is unity or slightly smaller. The final-state structure we obtain for T10F0.5 is shown in Figure \ref{fig:finalstate}.  In this limit a small amount of mass could conceivably be blown off as a wind, though this does not occur in our simulations, but it is not possible to eject enough matter to materially reduce the star formation efficiency. Bulk ejection appears likely only when the radiation field exerts enough force to be super-Eddington even using the lower opacity present at the top of the atmosphere. This condition is expected to be met under many circumstances, particularly in young star clusters \citep{thompson05a, krumholz09d, fall10a, murray10a}, and thus radiation pressure may still be a significant factor in limiting star formation efficiencies. However, models that rely on a large enhancement of the radiation force due to higher opacities in the warmer regions \citep[e.g.][]{murray10a, murray11a} may need to be reevaluated.  In particular, in the three unstable models considered here, the ratio of 
$\tau_{\rm IR}\simeq\kappa(T_{\rm mp})\Sigma$ to $f_{\rm trap}$ is $\simeq3-6$ (Table \ref{tab:simresult}).
Similarly, the models of \citet{krumholz09d} and \citet{fall10a} may also need to be reconsidered since they assume $\ft \sim 1$.
We find that $\ft$ can significantly exceed unity, which suggests that, if radiation does overcome gravity and eject matter, these models might underestimate the total momentum input to the gas by a factor as large as $\sim5-40$. However, since none of our runs successfully launch a wind, the question of how much momentum will be transferred to the gas if $\fEavg > 1$ and a wind is launched is not yet settled.

For subgrid models of radiative pressure feedback in numerical simulations, the main implication of our work is that the true value of $\ft$ is likely to be somewhat lower than $\tau_{\rm IR}$, and that $\ft$ will not increase in direct proportion to the total gas column density, or some proxy for it. Instead, we find that a reasonable estimate is that when $\fE > \fEc$, the system will adjust to give
\begin{equation}
\label{eq:ftrapest}
\ft \approx \frac{\tau_*}{\fE} - 1.
\end{equation}
Note that with this scaling $\ft$ still increases in direct proportion to $\Sigma$, as assumed in many models, but that there is an additional dependence on the flux, the opacity, and the gravitational force holding the gas together that ensures that the mean Eddington ratio $\fEavg$ saturates at unity. Simulations based on subgrid models in which $\ft$ is allowed to reach values such that $\fEavg$ is larger than unity need to be recomputed with lower values of $\ft$ to check which results are robust.

It is illuminating to compare our result to both the fiducial estimate of $\ft$ adopted by \citet{hopkins11a}, and to the alternative  models presented in Appendix B of their paper, where they attempt to take into account photon leakage (see also \citet{murray10a}). As noted above, in their fiducial models \citeauthor{hopkins11a} take $\ft = \tau_{\rm IR} - 1$. In the alternative models, they consider media with a variety of column density distributions motivated from observations of molecular clouds and simulations of turbulence. For the case that produces the most deviation from their fiducial estimate, that of an exponential distribution of column densities with a powerlaw tail, they find that for large optical depth the effective value of $\ft$ approaches
$\ft \approx \sqrt{\sigma/2\Gamma[1/\sigma]} \tau_{\rm IR}^{1/2\sigma} - 1$, where $\Gamma$ is the $\Gamma$ function, $\tau_{\rm IR}$ here is the optical depth that the medium would have if it were uniform, and $\sigma$ is the standard deviation of the column density distribution. \citeauthor{hopkins11a} consider values of $\sigma$ from $0.2 - 1.0$, and for the largest value of $\sigma$ they consider, at high $\tau_{\rm IR}$ this prescription therefore reduces to $\ft \approx \sqrt{\tau_{\rm IR}/2}-1$. Thus, for $\ft\gg 1$, the ratio of \citeauthor{hopkins11a}'s estimates of $\ft$ to the upper limit we measure in our simulations (Equation \ref{eq:ftrapest}) is
\begin{equation}
\label{eq:ftratio}
\frac{f_{\rm trap,Hopkins}}{f_{\rm trap,lim}} = \frac{\tau_{\rm IR}}{\tau_*} \fE\quad\mbox{or}\quad\sqrt{\frac{\tau_{\rm IR}}{2\tau_*^2}}\fE,
\end{equation}
where the first quantity is for \citeauthor{hopkins11a}'s fiducial estimate, and the second is for their alternative model with $\sigma = 1$. 
Similar estimates for $\ft$ in a turbulent medium were made by \citet{murray10a}).

\citeauthor{hopkins11a}'s fiducial estimate for $\tau_{\rm IR}$ is $\tau_{\rm IR} = 5 (\Sigma/\mbox{g cm}^{-2})$, whereas, for a normal galaxy in which $T_* \sim 40$ K, we will have $\tau_*\sim \tau_{\rm IR}/10$. Taking this as a rough estimate, Equation (\ref{eq:ftratio}) becomes
\begin{equation}
\label{eq:ftratio1}
\frac{f_{\rm trap,Hopkins}}{f_{\rm trap,lim}} \approx 10 \fE\quad\mbox{or}\quad\sqrt{\frac{\tau_{\rm IR}}{50}} \fE
\end{equation}
Thus we find that the \citet{hopkins11a} estimate of $\ft$ exceeds the limit we measure whenever $\fE > 0.1$ for their fiducial estimate, or when $\fE > \sqrt{50/\tau_{\rm IR}}$ for their lowest alternative estimate. The first condition is likely met in many places in their simulations. The second is somewhat harder to satisfy, but it still likely to be met at the highest optical depth locations in their simulations. 

The primary reason \citeauthor{hopkins11a}~and \citeauthor{murray10a}~estimate larger effective trapping
factors than we measure is that while their models consider the possibility that the gas might be non-uniform, they do not consider that the radiation might also be non-uniform, and that its non-uniformity might be correlated with that of the gas. Figures \ref{fig:flux} and \ref{fig:finalstate} show that, once the instability is fully established, this radiation-matter correlation is an essential feature of the radiation-gas coupling that these analytic models do not capture.

\subsection{Caveats and Future Work}
\label{section:caveats}

\subsubsection{Magnetic Fields}
\label{sec:bfields}

It is important to point out some potential limitations of our results, which suggest avenues for future investigation. The first is that we have omitted magnetic fields, which are clearly present in the real interstellar medium. Simulations of mechanical feedback from both protostellar outflows \citep{li06b, nakamura07a, wang10a} and photoionization-driven ``champagne" flows from molecular clouds \citep{gendelev12a} show that strong, ordered magnetic fields can significantly enhance the effects of feedback by providing a mechanism to transfer momentum between parcels of gas, and thus distribute energy and momentum more broadly through a gas cloud. This effect could conceivably operate here, and raise $\fEavg$ and $\ft$.

However, the effect is only likely to be important if the magnetic fields are dynamically strong; otherwise field lines will bend rather than exert significant forces. In ULIRGs, indirect arguments based on observations of synchotron emission suggest that the magnetic energy density is sub-equipartition compared to the turbulent or gravitational energies  \citep{thompson09a, lacki10a, batejat11a}, which suggests that magnetic effects are unlikely to be important there. The situation is less certain for massive star clusters in normal galaxies like the Milky Way. In star-forming clouds in the Milky Way and nearby galaxies, Zeeman and polarization observations indicate that there are ordered magnetic fields in rough equipartition with turbulence (\citealt{crutcher99a, troland08a, chapman11a, li11a}; however see \citealt{padoan04b} for a contrasting view), in which case magnetic effects could conceivably alter $\fEavg$ and $\ft$. However, all of the regions studied thus far are far from the regime of massive clusters forming at high volume and column density where radiative forces might be important, and it is unknown if the magnetic fields in such regions are also in equipartition. In any event, in future work it would be useful to investigate whether the inclusion of magnetic fields allows larger values of $\fEavg$ and $\ft$, and, if so, how large the field must be to achieve this effect.

\red{Another possible effect of magnetic fields is that they might make the gas subject to the local photon bubble instability of \citet{blaes03a}, as discussed in Section \ref{sec:nature}. In this case a local instability might exist on top of, or in place of, the interface one that occurs in our non-magnetic simulations. It seems unlikely that this additional instability would strengthen matter-radiation coupling, but it is conceivable that it could weaken it even further relative to what we have found.}

\subsubsection{Planar Versus Spherical Geometry}
\label{sec:planar}

For simplicity we have chosen to examine a planar geometry, since a spherical geometry would introduce a third parameter (which we could take to be the radius of curvature measured in units of $h_*$) to our description of the system. The main disadvantage of our planar approach is that in planar geometry the escape speed is not well-defined, and this precludes one possible mechanism for launching a wind. We find that, in steady state, $\fEavg$ reaches unity. However, there is a transient phase before the instability develops when $\fEavg$ is larger than unity. In planar geometry this makes little difference, because the momentum the gas absorbs during this transient is insufficient to escape, and the matter will always fall back. For a spherical geometry, however, it is conceivable that the gas could reach escape speed during the initial transient when $\fEavg>1$, allowing it to escape even though $\fEavg$ falls back to unity once steady state is reached.

This is unlikely to be a significant effect for the parameter regime we have explored. For galactic disks, non-spherical effects become significant only on scales comparable to the radial scale length, which is of order kpc. For comparison, for the ULIRG-like scalings shown in Table \ref{tab:simulations}, all our simulations boxes are $<1$ pc in size. Thus we can be confident that non-planar effects are unimportant in the galactic case unless the Eddington ratio, and thus the typical height to which matter is carried by radiation, is far larger than in the parameter regime we've explored. For the case of individual star clusters, which can be $\sim 1$ pc in size, the highest Eddington ratio case may be marginally in the regime where curvature effects become significant. However, even in this case most matter only reaches $\sim 0.3$ pc before turning around. Moreover, in a real star cluster where the initial matter distribution is not laminar, and where the radiation flux rises smoothly rather than turning on instantly as in our simulations, the amount of momentum deposited during the initial transient is likely to be less than in our simulations.

Thus we conclude that curvature effects are unimportant for the parameter regime we have explored, and are only likely to become important for systems with Eddington ratios significantly above 0.5.

\subsubsection{\red{2D versus 3D}}

\red{For reasons of computational efficiency we have conducted two-dimensional rather than fully three-dimensional simulations. While the dimensionality does not change the growth rate of the RRT instability in the linear regime \citep{jacquet11a}, it may affect the non-linear growth rate and fully saturated state of the instability. Even for simple fluid Rayleigh-Taylor instability the relationship between two- and three-dimensional results is still not fully understood, though numerical results indicate the non-linear growth rate is a factor of $\sim 2$ faster in the 3D case \citep[e.g.][and references therein]{young01a}. No comparable study exists for RRT, nor have we conducted such numerical experiments. However, it seems unlikely that the difference between 2D and 3D will be astrophysically important. The time required for the instability to reach full non-linear saturation in our simulations is smaller than essentially any other timescale relevant on galactic or star cluster scales, and a factor of order unity change in the non-linear growth rate would not alter this. Moreover, the non-linear saturated state we obtain is defined, as is the the case for many saturated instabilities, by the system self-regulating to a state of marginal stability, $\fEavg \approx 1$. It also seems unlikely that this self-regulation will fail in 3D. Thus the differences between 2D and 3D are unlikely to be astrophysically significant.
}

\subsubsection{\red{Dependence on Choice of Initial Conditions and Simulation Box Size}}

\red{Our computations use an idealized initial setup and computational box size chosen to allow us to follow the growth of the instability through its linear and then non-linear development, while ensuring that the smallest important size scale $h_*$ was always at least marginally resolved. It is fair to ask how the results might change for a more realistic setup that would correspond more closely to a real star cluster or ULIRG. In considering this question, it is helpful to separate the question of the initial horizontal and vertical structures.}

\red{In the horizontal direction, our initial conditions are characterized by a fairly small amplitude initial perturbation with power only at a single horizontal wavelength whose physical size is very small compared to real astrophysical systems -- for example, run T10F0.25 has only a 25\% density perturbation at a physical size of $0.082$ pc. Moreover, in the horizontal direction our entire computational box is only a factor of 2 larger than this. Since real protoclusters and ULIRGs are both supersonically turbulent and much larger than this, a real system would have much larger amplitude perturbations on much larger horizontal size scales. How would this affect our results?}

\red{In the linear regime, the analytic treatment discussed in Section \ref{sec:nature} is a helpful guide. The linear analysis shows that modes with short horizontal wavelengths are damped by horizontal diffusion of the radiation, which reduces the linear growth rate of the instability. Perturbations with longer horizontal wavelengths would be less damped, and thus should grow faster, with the growth timescale approaching $t_* \sim 1$ at long wavelengths, rather than $t_* \sim 10-100$ as found in our simulations.\footnote{\red{Conversely, shorter wavelength perturbations than those we have used would grow more slowly or not at all; indeed, in some low resolution simulations we conducted using even smaller horizontal box sizes, we did not see development of RRT instability. However, no real astrophysical system is this small horizontally.}} Moreover, with larger initial perturbations, fewer $e$-foldings would be required to reach the non-linear regime. Thus we conclude that in a real astrophysical system, the transition to the non-linear regime is likely to be significantly faster than in our simulations. Once in the non-linear regime, it seems likely that the same argument we made in considering 2D versus 3D applies: the non-linear state is characterized by the system self-regulating to a state of marginal stability, and it seems unlikely that this self-regulation depends on the size of the computational box or the initial state. Thus in the saturated state we do not expect the result that $\fEavg \approx 1$ would be different. However, the density structure might well extend to larger size scales in a larger computational domain. For ordinary fluid RT instability, in the non-linear regime the dominant mode is always on the largest scale permitted by the computational domain or experimental apparatus, and since RRT instability begins to behave like ordinary RT instability at very long wavelengths, this would likely be true for our problem too.}

\red{Our initial state is also very thin in the vertical direction, since it corresponds to a gas layer supported only by thermal pressure at constant temperature. In a real ULIRG or protocluster, the turbulent velocity greatly exceeds the thermal sound speed, and puffs up the gas to a much larger height, which is comparable to or larger than the vertical height we reach at the end of our simulations when the RRT instability is fully saturated. If we started with such a turbulent, vertically extended state but did not include any mechanism to drive the turbulence other than the RRT instability itself, then it seems likely that the result would be that the turbulence would decay until the gas reached the velocity dispersion dictated by the instability. The timescale for this to occur would likely simply be the turbulent decay time, which would be comparable to the crossing time in the initial vertical distribution. On the other hand, if there were some driving to maintain the turbulence, it is likely that this would dictate the gas structure. This probably could not cause $\fEavg$ to exceed unity, since this would imply that radiation rather than the other mechanism had become dominant. However, extra structure in the gas produced by turbulent driving could cause gas-radiation coupling to change in unexpected ways, and might conceivably lead to either stronger or weaker coupling of radiation and matter than would have occurred without the extra driving.}

\subsubsection{Nature of the Radiation Source}

Another caveat is that we have explored the effects of a radiation source that is steady, uniform, and located just below a gas layer, whereas in reality the young stars that provide radiation pressure in a young cluster or a ULIRG are spatially clustered, time-variable (as new stars form and old ones die), and are mixed with the gas. It seems highly unlikely that any of these effects would increase $\fEavg$ or $\ft$, since making the radiation field even less uniform would only further weaken matter-radiation coupling. On the other hand, as noted in Section \ref{sec:turborigin}, it seems possible that such non-uniform radiation source could produce larger turbulent velocities than we find in our simulations with a uniform radiation source. Again, followup simulations are needed to check this effect.

\subsubsection{\red{Numerical Treatment of the Radiative Transfer Problem}}

A final concern is \red{the quality of the two-temperature flux-limited diffusion (FLD) approximation we use to treat the radiative transfer problem. One significant concern is} our treatment of the direct (as opposed to dust-reprocessed) radiation field. The direct radiation field possesses two properties that are not well-captured by the FLD approximation. First, its spectrum is at much higher frequencies than would be predicted by our blackbody approximation; this is significant because dust opacities increase strongly with frequency. Second, the radiation field is highly directional, as opposed to the diffuse field assumed in the FLD approximation. In the case of the formation of a single massive star from an initially laminar protostellar core, \citet{kuiper11b} show that omitting the direct radiation field can lead to an underestimate of the expansion rate of radiation pressure-driven cavities, and that this in turn can affect whether a cavity goes RRT unstable before it blows out of its parent core.

This effect is unlikely to be important for the simulations we report here, simply because, by construction, the direct radiation pressure force is unimportant. The strength of the direct radiation pressure force relative to gravity can be characterized by the mean Eddington ratio considering only direct photons, which is given by Equation (\ref{eq:feavg}) evaluated with $\ft = 0$:
\begin{equation}
\fEdir = \frac{\fE}{\tau_*}.
\end{equation}
This gives $\fEdir = 0.002, 0.17, 0.025$, $0.05$, \red{and $0.09$} for runs T10F0.02, T03F0.50, T10F0.25, T10F0.50, \red{and T10F0.90} respectively. Thus we see that the direct radiation force is an order of magnitude or more weaker than gravity in our runs, and its omission is therefore not likely to produce significant effects.

It is important to note that the relative unimportance of the direct radiation field compared to gravity in our simulations is a direct result of our parameter choices, which are somewhat different from those appropriate to the single massive protostellar cores studied by \citet{kuiper11b}. To see why these systems are in a somewhat different regime than ULIRGs or star clusters, it is helpful to re-express $\fEdir$ in terms of dimensional parameters: $\fEdir = F_0 / \Sigma g c = L / M g c$, where in the second step we have multiplied through by a fiducial area to turn $F_0$ into a luminosity $L$ and $\Sigma$ into a mass $M$. If the gravitational field comes predominantly from the self-gravity of the object question, then $g\approx G \Sigma$, and we have
\begin{eqnarray}
\fEdir & \sim & f_* \frac{L/M_*}{G\Sigma c} 
\nonumber \\
&=&
4.8\times 10^{-4} \left(\frac{f_*}{1/2}\right) (L/M_*)_0 \Sigma_0^{-1},
\end{eqnarray}
where $(L/M_*)_0 = (L/M_*)/(\lsun/\msun)$, $\Sigma_0 = \Sigma/1$ g cm$^{-2}$, and for convenience we have defined $f_*$ as the stellar mass fraction, i.e.\ the fraction of the object's mass that is in luminous stars rather than gas. Note that this equation for the importance of direct radiation forces is, up to factors of order unity, the same as the result derived by \citet[their Equation 5]{fall10a}. Individual massive protostellar cores, massive protoclusters, giant clumps in high redshift galaxies, and ULIRG disks all have similar surface densities $\Sigma \approx 1$ g cm$^{-2}$, but they have very different stellar light to mass ratios. For example, a 100 $\msun$ star on the zero-age main sequence (ZAMS) has $L/M_* = 1.3\times 10^4 \lsun/\msun$ \citep{tout96a}, so $\fEdir$ can easily be significantly greater than unity, particularly in a rotating system where rotational flattening lowers the surface density in certain directions. In contrast, a cluster of 
ZAMS stars drawn from a fully sampled IMF will have $L/M_* \approx 10^3 \lsun/\msun$ \citep{murray10b}, an order of magnitude lower.
Since $L/M_*$ declines as a stellar population ages, $\fEdir$ will likely be smaller than unity in ULIRG disks or in the giant clumps in high redshift galaxies with dynamical times longer than $\sim 4$ Myr \citep{krumholz10b}, although $\fEdir$ might still exceed unity in subregions where the stellar population is predominantly young. We can therefore conclude that, even if the direct radiation field can be dominant for single massive stars and cores, it is at most an order unity effect for massive star clusters and ULIRGs (but see \citet{murray10a} and \citet{krumholz10b} for more discussion).

Furthermore, as \citet{socrates08a} point out, the above calculation may overestimate the importance of the direct radiation force if the geometry of the emitters is uniform, because it does not consider the fact that the direct radiation forces supplied by stars can cancel. A single massive star supplies a radially outward radiation field, but the many stars in a cluster or disk will push the gas with which they are intermingled in many different directions, and thus there will be some cancellation, with the exact amount depending on the geometry.
However, note that \citet{hopkins11a} argue that the effects of cancellation are not strong in 
simulations of radiation pressure feedback in simulations of whole galaxies.

\red{Even if omission of the direct radiation field is not problematic in our physical situation, one can still worry about
other aspects of the FLD approximation. FLD requires that the region of interest be optically thick, and our simulation domain certainly satisfies that criterion: $\tau_{\rm IR}$ exceeds 10 in all the runs and exceeds 100 in three of them (see Table \ref{tab:simresult}). However, since FLD discards information on the directionality of the radiation field, it underestimates beaming of the radiation field in localized patches that may be optically thin. Quantifying the net effect of this error is difficult, and generally requires direct comparison of FLD results with more accurate methods such as variable Eddington tensor, discrete ordinates, or Monte Carlo. One such comparison, done in the context of a radiation-dominated accretion disk by \citet{davis12a}, suggests that FLD tends to overestimate the vertical force exerted by the radiation, because it underestimates the ability of the radiation to stream in the horizontal direction. If this were true of our problem it would suggest that we have overestimated $\ft$. However, we caution that \citeauthor{davis12a}'s problem is not completely analogous to ours (for example the dominant opacity for \citeauthor{davis12a}~is the scattering opacity of free electrons, rather than the absorption opacity of dust), and it is not clear if the same would be true for our case.
}

\section{Conclusions}
\label{sec:summary}

We present numerical simulations of a strong radiation flux passing through a column of gas confined by gravity. This configuration is a reasonable proxy for a galactic disk or a section of a young star cluster in which the radiation from young, massive stars passes through the dusty interstellar medium and exerts dynamically significant forces. This system is characterized by two dimensionless numbers, $\tau_*$ and $\fE$. The former describes the optical depth of the gas column computed using the temperature at its surface, and the latter describes the ratio of radiation pressure force to gravitational force in this gas. We use these simulations to study whether radiation is able to drive turbulence or produce winds in the regime where the radiation force is sub-Eddington for the cool gas at the top of the disk / the edge of the young cluster ($\fE<1$), but the gas is optically thick ($\tau_* > 1$), so that the higher temperatures make the gas super-Eddington near the disk midplane / cluster center. The disks of ULIRGs and some young star clusters in non-ULIRG galaxies are in this regime.

We find that, in this regime, radiation forces drive the matter into a thin sheet which then breaks up due to radiation Rayleigh-Taylor instability (Figs.~\ref{fig:rhoplot}, \ref{fig:flux}, \ref{fig:finalstate}). Once this instability reaches its full non-linear development, it taps a fraction of the radiation energy and uses it to drive supersonic turbulent motions in the gas. In the parameter regime appropriate to ULIRGs and young cluster clusters the velocity dispersion can approach Mach 10, sufficient to fully explain the turbulence observed in young protocluster gas clouds in the Milky Way. ULIRGs show significantly greater velocity dispersions, which suggests that either 
radiation pressure-driven instabilities cannot drive turbulence at the required levels, or that ULIRGs are closer to 
the Eddington limit at the cool tops of their atmospheres than current observations suggest; this is entirely possible, given the observational uncertainties.

We also find that radiation Rayleigh-Taylor instability leads to a configuration in which the matter is concentrated in filaments, while the radiation flux is concentrated in low-density channels (Figs.~\ref{fig:flux}, \ref{fig:finalstate}). In this configuration radiation is not fully trapped by the gas. As a result, the actual mean force applied to the gas never exceeds that applied by gravity, despite the fact that gas near the midplane is super-Eddington. We find radiation passing through slabs of matter with $\fE<1$ does not drive significant mass ejection or a noticeable wind even if $\tau_* \fE \gg 1$. The ratio of the momentum transferred to the gas to that carried by the radiation field, defined as $1+\ft$ where $\ft$ is the trapping factor, reaches a value (given by Equation \ref{eq:ftrapest}) that ensures that the mean Eddington ratio is unity. These numerical results are in conflict with the assumptions built into analytic models and sub-grid implementations of radiation pressure feedback in numerical simulations, which either limit $\ft\sim1$ or allow substantially larger values of $\ft$. Simulations based on these assumptions should be recomputed using our improved determination of $\ft$.

\begin{figure}
\plotone{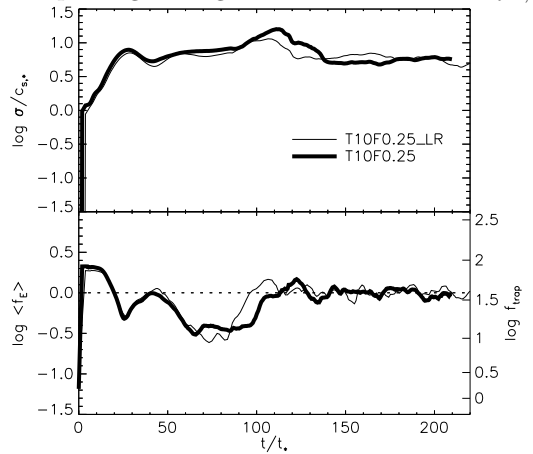}
\caption{
\label{fig:resstudy}
\red{Comparison of the total velocity dispersion $\sigma$ (top panel), mean Eddington ratio $\fEavg$ (bottom panel, left axis), and trapping factor $\ft$ (right axis) in runs T10F0.25 (thick lines) and T10F0.25\_LR (thin lines). The results are clearly quite similar. To avoid clutter, we do not show $\sigma_x$, $\sigma_z$, or the trapping factor considering only material with $v_z > 0$, as we do in Figures \ref{fig:vdisp} and \ref{fig:frad}, but these lines are also very similar in the two runs.}
}
\end{figure}

\begin{appendix}

\section{\red{Resolution Study}}

\red{To check the dependence of our results on the numerical resolution of the simulations, we rerun simulation T10F0.25 at half resolution; we call this run T10F0.25\_LR, and describe its properties in Table \ref{tab:simnum}. Since the instability in its fully saturated state is chaotic, we do not expect the results of runs T10F0.25 and T10F0.25\_LR to be identical in more than a statistical sense in the non-linear regime. Figure \ref{fig:resstudy} shows a comparison of the time evolution of the gas velocity dispersion, Eddington ratio, and trapping factor in the two runs, and Table \ref{tab:simresult} gives the quantitative results. As the Figures and Table show, the results are in line with what one would expect for a converged calculation. At early times, before the instability becomes non-linear, $\sigma$, $\fEavg$, and $\ft$ are nearly identical in the two runs. As time goes on, the instability goes non-linear, and the flow becomes chaotic, the two runs diverge, but they remain statistically nearly identical. In particular, the primary result that the mean Eddington ratio self-regulates to unity at late times once the instability is fully saturated is found at both resolutions. The time-averaged values of $\sigma$ and $\tau_{\rm IR}$ differ by $\sim 10\%$ between the two runs, but these differences are smaller than the fluctuations in time in these quantities found in each run, and thus are consistent with simply being the results of random sampling of the chaotic flow pattern.
}

\end{appendix}

\acknowledgements MRK acknowledges support from the Alfred P.~Sloan Foundation,  the NSF through grant CAREER-0955300, and NASA through Astrophysics Theory and Fundamental Physics Grant NNX09AK31G, and a Chandra Space Telescope Grant.
TAT acknowledges support from the Alfred P.~Sloan Foundation and NASA grant NNX10AD01G.


\end{document}